\documentclass[]{aa}
\usepackage{graphics}
\begin{document}

\newcommand{\Tcool}{\mbox{$T_{\mathrm {cool}}$}}
\newcommand{\Rcool}{\mbox{$R_{\mathrm {cool}}$}}
\newcommand{\Ncool}{\mbox{$N_{\mathrm {cool}}$}}
\newcommand{\Thot}{\mbox{$T_{\mathrm {hot}}$}}
\newcommand{\Rhot}{\mbox{$R_{\mathrm {hot}}$}}
\newcommand{\Nhot}{\mbox{$N_{\mathrm {hot}}$}}
\newcommand{\Vmax}{\mbox{$\mathrm{V}_{\mathrm {max}}$}}
\newcommand{\Vmin}{\mbox{$\mathrm{V}_{\mathrm {min}}$}}
\newcommand{\Rin}{\mbox{$R_{\mathrm {in}}$}}
\newcommand{\Rout}{\mbox{$R_{\mathrm {out}}$}}

\title{The time variation in infrared water-vapour bands in Mira variables
      \thanks{
      Based on observations with ISO, an ESA project with instruments
      funded by ESA Member States (especially the PI countries: France,
      Germany, the Netherlands and the United Kingdom) 
      with the participation of ISAS and NASA. 
      The SWS is a joint project of SRON and MPE.}
}
\author{ M.\,Matsuura       \inst{1,2,3}
   \and  I.\,Yamamura       \inst{1}
   \and  J.\,Cami           \inst{4,5}
   \and  T.\,Onaka          \inst{2}
   \and  H.\,Murakami       \inst{1}
}
\institute{
      The Institute of Space and Astronautical Science (ISAS),
      Yoshino-dai 3-1-1, Sagamihara, Kanagawa 229-8510, Japan
   \and
      Department of Astronomy, School of Science, University of Tokyo,
      Hongo 7-3-1, Bunkyo, Tokyo 113-0033, Japan
  \and
	Department of Physics, UMIST, P.O. Box 88, Manchester M60 1QD, UK
   \and
      Astronomical Institute `Anton Pannekoek', University of Amsterdam,
      Kruislaan 403, 1098 SJ, Amsterdam, the Netherlands
   \and
      SRON-Groningen, P.O. Box 800, 9700 AV Groningen, the Netherlands
}
\abstract{The time variation in the water-vapour bands
in oxygen-rich Mira variables has been investigated using
multi-epoch ISO/SWS spectra of four Mira variables
in the 2.5--4.0\,$\mu$m region.
All four stars show H$_2$O bands in absorption
around minimum in the visual light curve.
At maximum, H$_2$O emission features appear
in the $\sim$3.5--4.0\,$\mu$m region,
while the features at shorter wavelengths remain in absorption.
These H$_2$O bands in the 2.5--4.0\,$\mu$m region originate from the extended atmosphere.

The analysis has been carried out with
a disk shape, slab geometry model.
The observed H$_2$O bands are reproduced by two layers;
a `hot' layer with an excitation temperature of 2000\,K
and a `cool' layer with an excitation temperature of 1000--1400\,K.
The column densities of the `hot' layer are
$6\times10^{20}$--$3\times10^{22}$\,cm$^{-2}$,
and exceed $3\times10^{21}$\,cm$^{-2}$
when the features are observed in emission.
The radii of the `hot' layer (\Rhot) 
are $\sim$1\,$R_*$ at visual minimum 
and 2\,$R_*$ at maximum,
where $R_*$ is a radius of background source of the model,
in practical, the radius of a 3000\,K black body.
The `cool' layer has the column density (\Ncool) 
of $7\times10^{20}$--$5\times10^{22}$\,cm$^{-2}$,
and is located at 2.5--4.0\,$R_*$.
\Ncool\ depends on the object rather than the variability phase.

The time variation of $\Rhot/R_*$ from 1 to 2 is attributed to
the actual variation in the radius of the H$_2$O layer,
since the variation in \Rhot\ far exceeds the variation 
in the `continuum' stellar radius.
A high H$_2$O density shell occurs near the surface of the star around minimum,
and moves out with the stellar pulsation.
This shell gradually fades away after maximum,
and a new high H$_2$O density
shell is formed in the inner region again at the next minimum.
Due to large optical depth of H$_2$O,
the near-infrared variability is dominated by the H$_2$O layer,
and the L'-band flux correlates with the area of the H$_2$O shell.
The infrared molecular bands trace
the structure of the extended atmosphere
and impose appreciable effects on near-infrared light curve of Mira variables.
}
\titlerunning{The time variation in water-vapour bands in Mira}
\authorrunning{M.\,Matsuura et al.}
\date{Received ; Accepted}
\offprints{M.M. (m.matsuura@umist.ac.uk)}

\maketitle

\section{Introduction}
 Asymptotic Giant Branch (AGB) stars 
are in the late stage of the stellar evolution
for low and intermediate main-sequence mass stars.
Generally, AGB stars are pulsating variables as represented by Mira variables.
According to hydrodynamic model atmospheres,
the pulsations lift up matter from the stellar surface
and extend the atmosphere
(e.g. Bowen~\cite{Bowen88}).
Pulsations create shocks,
causing a step-like structure in the density
distribution as a function of radius
(e.g. Fleischer et al.~\cite{Fleischer92};
H\"ofner et al.~\cite{Hoefner98}).
The cooling behind the shock is efficient
(Woitke et al.~\cite{Woitke96})
and 
the temperature decreases immediately in the post-shock regions
(Fleischer et al.~\cite{Fleischer92};
H\"ofner et al.~\cite{Hoefner98}).

 The extended atmosphere is filled with various kinds of molecules.
The structure of the extended atmosphere can be studied using
infrared molecular bands.
Hinkle (\cite{Hinkle78}) and Hinkle \& Barnes (\cite{Hinkle79})
analyzed high resolution spectra of the oxygen-rich Mira, \object{R~Leo},
and found two velocity components in the near-infrared
molecular lines of CO, OH, and H$_2$O.
They concluded that
one component is located near the boundary region of the photosphere
and the second component is superposed on the first layer
above the photosphere.

However, the molecular bands suffer interference from
molecules in the terrestrial atmosphere. 
Recent space-borne observations enable more comprehensive
studies of the molecules
in the extended atmosphere.
Using 
the Short-Wavelength Spectrometer (SWS; de~Graauw et al.~\cite{deGraauw96})
on board the Infrared Space Observatory (ISO; Kessler et al.~\cite{Kessler96}),
Tsuji et al. (\cite{Tsuji97}) found
CO, H$_2$O, CO$_2$ and SiO molecules located above the photosphere.
Markwick \& Millar (\cite{Markwick00})
indicated that the 2.8\,$\mu$m spectra of a Mira variable
consists of two H$_2$O components with 950\,K and 250\,K.
Yamamura et al. (\cite{Yamamura99a}) identified SO$_2$ features
in the 7\,$\mu$m region in three oxygen-rich Mira variables
with an excitation temperature estimated to be 600\,K.
The band was found to be variable, changing from emission to absorption.
The time scale of the SO$_2$ variation was longer than 
the period of the visual variable phase.
Justtanont et al. (\cite{Justtanont98}) and
Ryde et al. (\cite{Ryde99a})
detected CO$_2$ bands in the 12--17\,$\mu$m region.
Cami et al. (\cite{Cami00}) 
found these CO$_2$ molecules fill the region between 4--400\,stellar radii.
Not only molecules but also fine-structure atomic lines
were found in the upper atmosphere (Aoki et al.~\cite{Aoki98a}).
 
H$_2$O is one of the most abundant molecules in the oxygen-rich
atmosphere and is a large opacity source in the near-infrared region.
Tsuji (\cite{Tsuji78b}), using low resolution spectra obtained 
by the Kuiper Airborne Observatory,
suggested that the 5--8\,$\mu$m region is filled with the H$_2$O emission
arising from the atmosphere above the photosphere
(or the hydrostatic atmosphere).
However, the spectral resolution
of those data was too low for further study.
Yamamura et al. (\cite{Yamamura99b}) found
emission features from water-vapour bands around 3.5--4.0\,$\mu$m in $o$~Cet,
which was observed at maximum in the visual light curve.
In contrast, \object{Z~Cas}, which was observed near minimum,
showed absorption features at the same wavelengths.
They analyzed water-vapour spectra from 2.5 to 4.0\,$\mu$m
with a simple `slab' model.
The model consists of two molecular layers
(`hot' layer and `cool' layer)
with independent excitation temperatures, 
column densities, and radii.
The `hot' layer with an excitation temperature of 2000\,K
extended to $\sim$2\,$R_*$ in $o$~Cet and stayed
at $\sim$1\,$R_*$ in Z~Cas,
where $R_*$ is the radius of the background light source representing
the star.
They surmised
that water layers are generally more extended at maximum.

In this paper we examine the time variation in the water vapour bands
in the 2.5--4.0\,$\mu$m region.
Yamamura et al. (\cite{Yamamura99b}) analyzed two different stars
at different phases.
To investigate whether the difference in the radii
of the water layers is related to the phase difference or not,
we analyzed the spectra of four Mira variables observed several
times with the ISO/SWS.
We found periodical variation in H$_2$O bands in ISO/SWS spectra.
Some features turn from absorption to emission during minimum and maximum.
This variation is explained by the variation in the radius
of the H$_2$O layer. These H$_2$O molecules are located in the extended atmosphere.
We discuss the variation in the structure
of the extended atmosphere caused by the pulsations.

\section{Observational data}

 We extracted SWS observations from the ISO Data Archive.
In Table~\ref{table-gcvs}
we summarize the sampled stars and their properties,
and in Table~\ref{table-observation} we show
the observation journal.
These data are collected from three SWS programs:
AGBSTARS (P.I. T.~de~Jong),
TIMVAR (P.I. T.~Onaka), 
VARLPV (P.I. J.~Hron).
The data and results of 
\object{$o$~Cet} and Z~Cas reported in Yamamura et al. (\cite{Yamamura99b})
are also listed.
Four stars (\object{R~Aql}, \object{R~Cas}, \object{T~Cep}, and \object{Z~Cyg}) 
were observed from 5 to 7 times and the observed 
time span covered more than
one variability period of each star
(Onaka et al.~\cite{Onaka99}; Loidl et al.~\cite{Loidl99}).
These data are ideal to examine the time variation in the spectra.
The spectra were obtained using the full-grating scan mode
(AOT~01, scan speed is 1, 2, and 3).
The wavelength coverage was 2.35--45.2\,$\mu$m.
The spectral resolution was $\lambda / \Delta \lambda =300$--1000,
depending on the wavelength and the scan speed.
The data were reduced using the SWS Interactive Analysis package
(e.g. Wieprecht et al.~\cite{Wieprecht01}).
The calibration parameters at October 1999
are used for the wavelength, detector responsivity,
and absolute flux calibrations.
%
%
The photometric accuracy of the observed spectra is
5--7\,\% for bright sources in band-1, i.e. below 4\,$\mu$m
(Decin et al.~\cite{Decin00}).
Small differences in the flux level between different
AOT bands were corrected by scaling
the flux at each band with respect to the flux in
band-1b (2.6--3.0\,$\mu$m).
The correction factor is within a few percent in
band-1.
%
%
The spectra are re-gridded with a constant wavelength resolution
of $\lambda / \Delta \lambda =300$, assuming the Gaussian
wavelength profile at each wavelength grid. The re-gridded
spectra are oversampled by a factor of five.

\begin{table}
\begin{caption}
{The list of target Mira variables. 
The properties are taken from the General Catalogue
of Variable Stars
(GCVS; Kholopov et al.~\cite{Kholopov88}),
and 
are estimated from the light curve within two years obtained by 
AAVSO (Fig.~\ref{aavso}; Mattei~\cite{Mattei99}).
}\label{table-gcvs}
\end{caption}
\begin{tabular}{r ll r@{.}l lll} \hline
Name &\multicolumn{4}{l}{GCVS} & \multicolumn{3}{l}{AAVSO}\\
& \Vmax & \Vmin &\multicolumn{2}{c}{Period} & \Vmax & \Vmin  & Period \\
        & [mag] & [mag] & \multicolumn{2}{c}{[days]} & [mag] & [mag] & [days] \\ \hline   
R~Aql		&5.5	&12.0	&284	&2	& 6.5	& 10.8	&274	\\
R~Cas		& 4.7	&13.5	&430	&46	& 6.5	&12.0	&444	\\
T~Cep		& 5.2	&11.3	&388	&14	& 6.1	&10.6	&402	\\
Z~Cyg		& 7.1	&14.7	&263	&69	& 8.9	&13.4	&265	\\\hline
Z~Cas		& 8.5	&15.4	&495	&71	&&&\\
$o$~Cet	& 2.0	&10.1	&331	&96	&&&\\\hline
\end{tabular}
\end{table}

\begin{table}
\begin{caption}
{Journal of the ISO/SWS observations.
The phases are estimated from the optical light curves provided by
the AAVSO (Mattei~\cite{Mattei99}).
$F_{\nu}$ at L'-band 
is calculated from filter transmission of
ESO L'-band (effective wavelength is 3.771\,$\mu$m;
van der Bliek et al.~\cite{vanderBliek96}).
The last column is for the ISO program name.
}\label{table-observation}
\end{caption}
\begin{tabular}{lrcrr@{.}l c} \hline

Name		&  \multicolumn{1}{c}{Observed}	& Scan		& Phase		& \multicolumn{2}{c}{$F_{\nu}$ at}		 & Pro. \\
		&  \multicolumn{1}{c}{date}		& speed		&		& \multicolumn{2}{c}{L' [Jy]}		& 	\\ \hline
R~Aql-1		& Mar. 18, 1996		& 1 	&$-0.08$	&  778&3 & 1\\
R~Aql-2		& Oct. 01, 1996		& 2 	& 0.64		&  820&4 & 1\\
R~Aql-3		& Mar. 03, 1997		& 2 	& 1.21		& 1051&1 & 1\\
R~Aql-4		& May  03, 1997		& 2 	& 1.42		&  811&9 & 1\\
R~Aql-5		& Sep. 21, 1997		& 2 	& 1.99		&  741&1 & 1\\
R~Cas-1		& July 21, 1996		& 2 	& 0.46		& 2156&4 & 1\\
R~Cas		& Dec. 03, 1996		& 2 	& 0.78		& 2372&9 & 2\\
R~Cas-2		& Dec. 03, 1996		& 2 	& 0.78		& 2322&1 & 1\\
R~Cas-3		& Dec. 15, 1996		& 2 	& 0.80		& 2324&3 & 1\\
R~Cas-4		& Jan. 10, 1997		& 2 	& 0.86		& 2367&5 & 1\\
R~Cas-5		& Feb. 02, 1997		& 2 	& 0.92		& 2532&4 & 1\\
R~Cas-6		& Aug. 04, 1997		& 2 	& 1.35		& 2849&4 & 1\\ 
T~Cep-1		& Aug. 03, 1997		& 1 	& 0.38		& 2258&7 & 3   \\
T~Cep-2		& Oct. 27, 1996		& 2 	& 0.58		& 1909&8 & 3   \\
T~Cep-3		& Jan. 15, 1997		& 2 	& 0.79		& 1832&4 & 3   \\
T~Cep-4		& Apr. 13, 1997		& 2 	& 1.02		& 2127&7 & 3   \\
T~Cep-5		& June 13, 1997		& 2 	& 1.17		& 2528&6 & 3   \\
T~Cep-6		& Sep. 07, 1997		& 2 	& 1.39		& 2148&1 & 3   \\
T~Cep-7		& Dec. 01, 1997		& 2 	& 1.61		& 1964&0 & 3   \\
Z~Cyg-1	& Aug. 05, 1996		& 1 	& 0.55		&	44&8 & 3   \\
Z~Cyg-2	& Oct. 08, 1996		& 2 	& 0.79		&	34&5 & 3   \\
Z~Cyg-3	& Nov. 24, 1996		& 2 	& 0.97		&	55&4 & 3   \\
Z~Cyg-4	& Jan. 24, 1997		& 2 	& 1.20		&	65&8 & 3   \\
Z~Cyg-5	& Mar. 21, 1997		& 2 	& 1.42		&	55&6 & 3   \\
Z~Cyg-6	& May~  15, 1997	& 2 	& 1.63		&	48&2 & 3   \\\hline
$o$~Cet		& Feb. 09, 1997		& 3 	& 0.99		& 5773&5 & 2 \\
Z~Cas		& Feb. 26, 1996		& 2 	& 0.50		&  148&9 & 2 \\\hline
\end{tabular}

Program name 1: VARLPV (P.I. J.~Hron); 2: AGBSTARS (P.I. T.~de~Jong);
3: TIMVAR (P.I. T.~Onaka).
\end{table}

\begin{figure}[!ht]
\resizebox{\hsize}{!}{\includegraphics*[60,320][560,660]{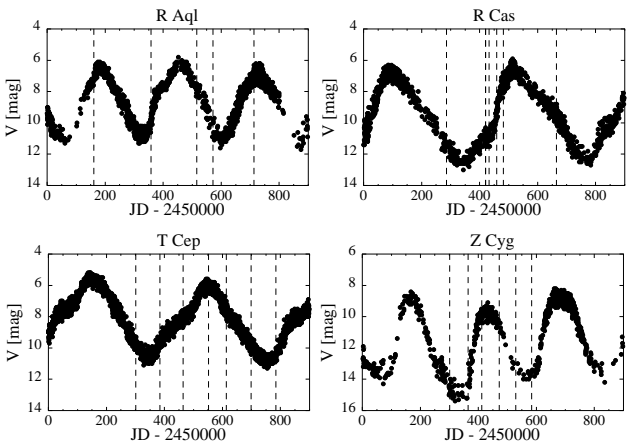}}
\caption{
The light curves obtained by AAVSO (Mattei~\cite{Mattei99}).
The epochs of the ISO observations are denoted as dotted lines.
}
   \label{aavso}
\end{figure}

\subsection{An overview of the observed spectra}
\begin{figure}[!ht]
\begin{center}
\resizebox{\hsize}{!}{\includegraphics*[90,373][555,720]{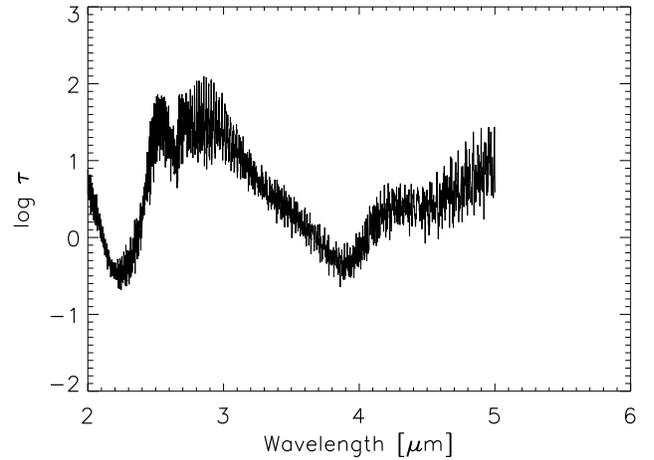}}
\end{center}
\caption{
$^1$H$_2^{16}$O optical depth ($\tau$) calculated 
from the line list of Partridge \& Schwenke (\cite{Partridge97}).
Parameters are a line width of 5\,km\,s$^{-1}$,
a column density of $3\times10^{21}$\,cm$^{-2}$ multiplied 
by isotopic abundance ratio (0.997),
and an excitation temperature of 2000\,K.
H$_2$O lines continuously occupy the spectrum in the 2.5--5.0\,$\mu$m region.
These lines are mainly the vibrational transitions
of $\Delta v_1=1$, $\Delta v_2=2$, 
and $\Delta v_3=1$
(Table~\ref{table-H2O lines}).
(Author note: This figure is modified from the original A\&A version,
 due to the restriction of file size of astro-ph.
 The optical depth is smoothed with a resolution of R=2000.)
}
   \label{fig-tau}
\end{figure}

\begin{figure*}[!ht]
\begin{center}
\resizebox{\hsize}{!}{\includegraphics*[35,270][540,740]{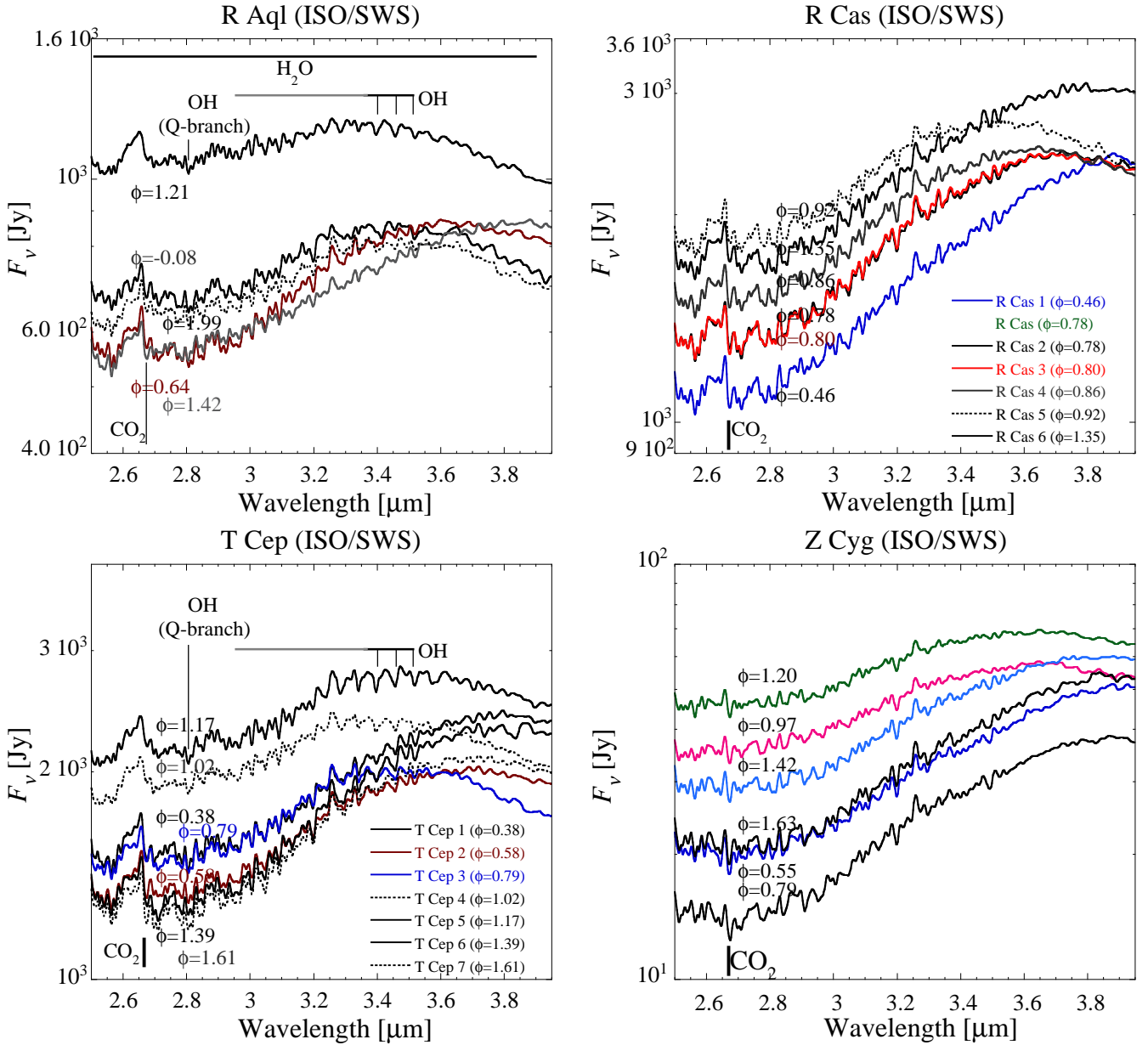}}
\end{center}
\caption{
ISO/SWS spectra.
The optical variability phase ($\phi$),
which is estimated from AAVSO light curve (Mattei~\cite{Mattei99})
at each observation, is indicated by the labels.
The 2.7\,$\mu$m absorption with respect to the
flux maximum around 3.5\,$\mu$m
is generally shallower around maximum than minimum.
The flux levels return to a comparable level
after one period.
There are two spectra for $\phi$=0.78 (on the same day)
for R~Cas,
and only R~Cas in the program of VARLPV is plotted.
}
   \label{obs}
\end{figure*}

The observed spectra are shown in Fig.~\ref{obs}.
The positions of the major molecular bands are indicated.
Due to their large optical depth (Fig.~\ref{fig-tau}),
H$_2$O features dominate
both the global shape and the small features
between 2.5 and 4.0\,$\mu$m.
CO$_2$ bands are visible at 2.67\,$\mu$m.
OH absorption features are seen around 3.5\,$\mu$m
in R~Aql and T~Cep, especially near maximum, 
while these features are not clear in R~Cas and Z~Cyg.

All of the four stars indicate periodical variations.
The 2.7\,$\mu$m absorption relative to the maximum
flux around 3.5\,$\mu$m
is generally shallower around maximum than minimum.
The flux levels return to a comparable level
after one period,
for example, in T~Cep ($\phi$=0.58, 1.39, and 1.61)
and in Z~Cyg ($\phi$=0.55, and 1.63).

\begin{figure}[!ht]
\begin{center}
\resizebox{\hsize}{!}{\includegraphics*[40,120][500,655]{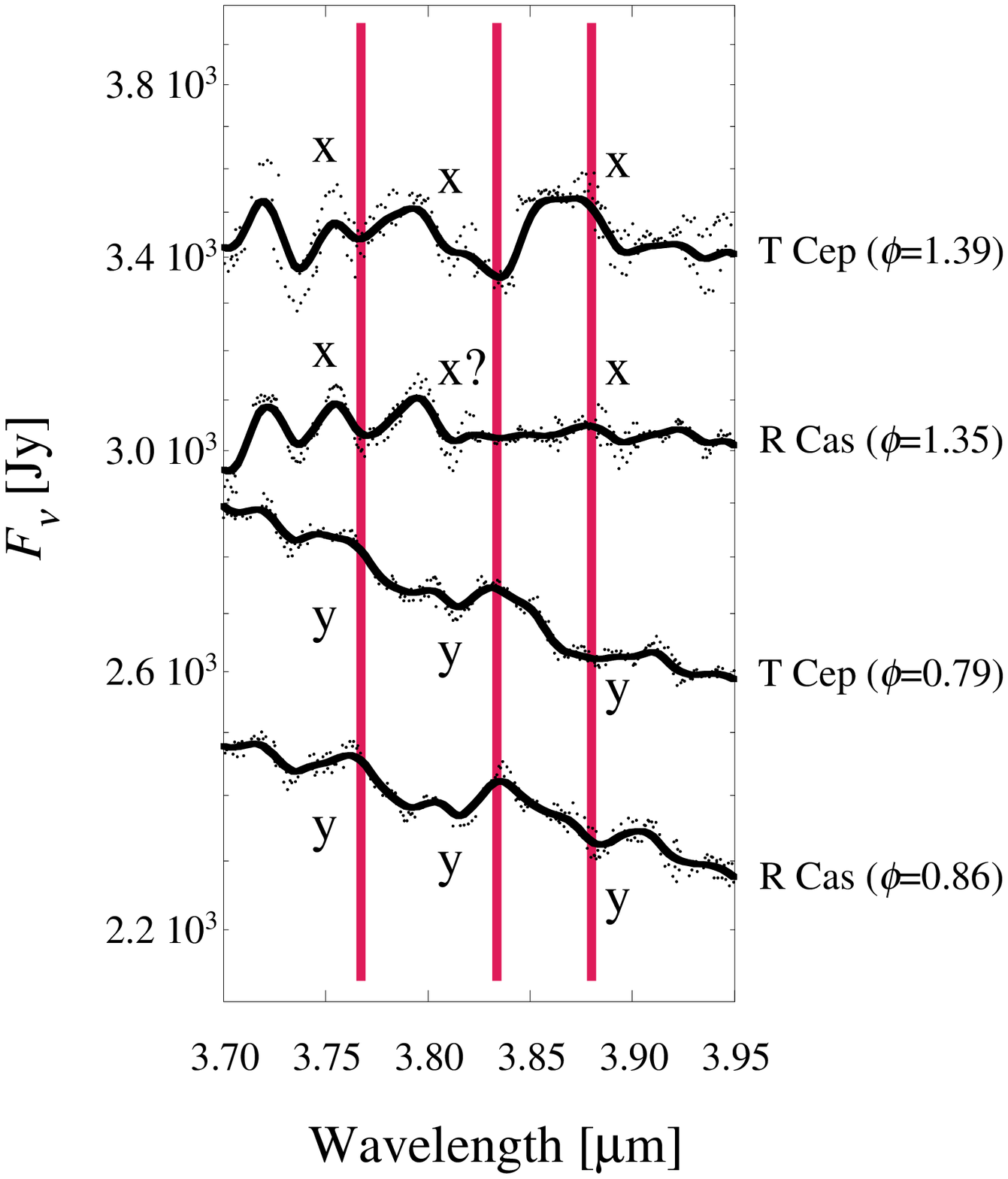}}
\end{center}
\caption{
Representative ISO/SWS spectra in the 3.8\,$\mu$m region.
Three conspicuous features, which show time variation,
are marked as `x' and `y'.
These features are inverted between the spectra of
T~Cep ($\phi$=1.39) and R~Cas ($\phi$=1.35)
and the ones of T~Cep ($\phi$=0.79) and R~Cas ($\phi$=0.86).
As we discuss in Sect.~\ref{sec-radii},
`x' is an absorption phase and `y' is an emission phase of the feature.
The feature at 3.88\,$\mu$m is opposite to its aspect.
The flux of T~Cep is scaled by a factor of 1.5.
The dots are ISO/SWS data sampling points before re-gridding
to constant wavelength grid.
Although there is a scatter in the data presented,
we could demonstrate that
the phases of the features are correct
and height of the features generally have uncertainty of up to 10\,\%.
}
   \label{spec_3.8_asc}
\end{figure}

 Representative ISO/SWS spectra in the 3.8\,$\mu$m region
are shown in Fig.~\ref{spec_3.8_asc}.
There are three conspicuous features which show time variation.
The appearance of these features is inverse from minimum to maximum.
As we will discuss in Sect.~\ref{sec-radii},
the features denoted with `x' are in absorption
and the ones with `y' are in emission.
The definition of the absorption or emission at 3.88\,$\mu$m
is opposite to the aspect.
%

\section{Model analysis}

\subsection{Two-layer `slab' model}

Water spectra in Mira variables are well represented 
by a model consisting of 
two water layers
(Yamamura et al.~\cite{Yamamura99b}).
In $o$~Cet, one layer contributes to the emission
features longer than $\sim$3.5\,$\mu$m,
and the other layer contributes to the absorption 
features shorter than $\sim$3.5\,$\mu$m.
These two layers are called `hot' layer and `cool' layer, respectively.

\begin{figure}[!ht]
\begin{center}
\resizebox{\hsize}{!}{\includegraphics*[5,140][510,710]{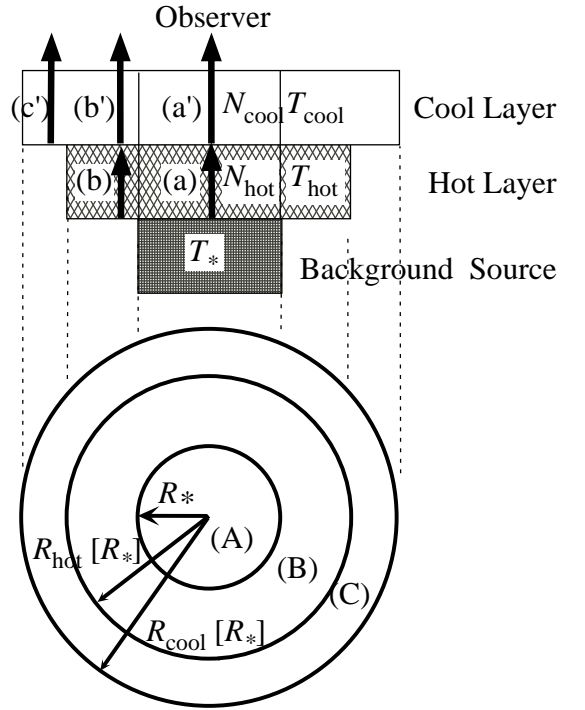}}
\end{center}
\caption{
Model configuration. Two disk-shape molecular layers are superposed
on a background source,
and the radiative transfer is solved though the layers.
The top panel indicates the side view
and the bottom panel indicates
the geometry seen by the observer.
}
   \label{config}
\end{figure}

We use the same model as Yamamura et al. (\cite{Yamamura99b}).
Plane-parallel, disk shape (or circular slabs), 
uniform molecular layers are superposed,
and the radiative transfer through the layers is calculated.
A schematic view of this model configuration is indicated in Fig.~\ref{config}.
Each layer is described by three parameters,
the excitation temperature (\Thot\ or \Tcool),
the column density (\Nhot\ or \Ncool), and 
the radius of the layer (\Rhot\ or \Rcool).
The radii of the layers are given  relative to $R_*$,
the radius of the background continuum source.

 The `hot' layer has an excitation temperature of about 2000\,K.
The water spectrum longer than $\sim$3.5\,$\mu$m
is dominated by this `hot' layer
(seen in Fig.~\ref{Rh}, for example).
The H$_2$O lines in this wavelength region are mainly caused
by higher excitation levels of $\Delta v_2=2$,
$\Delta v_1=1$, and $\Delta v_3=1$.
The upper `cool' layer has an excitation temperature
of 1000--1400\,K 
and is seen below $\sim$3.5\,$\mu$m,
where relatively low excitation levels
of $\Delta v_1=1$ and $\Delta v_3=1$ dominate.
This `cool' layer becomes optically thick 
below $\sim$3.5\,$\mu$m
for a column density of 
$1\times 10^{20}$--$1\times 10^{21}$\,cm$^{-2}$
 (Table~\ref{table-H2O tau}).
The column densities in Mira variables usually exceed these values
(Yamamura et al.~\cite{Yamamura99b};
summarized in Table~\ref{table-yamamura}).
Thus, the features from the `hot' layer are masked
by the `cool' layer below $\sim$3.5\,$\mu$m.

The `disk shape' plane-parallel model (Fig.~\ref{config})
is a simplification of the spherical geometry.
This simplification
enables us to treat more than one million H$_2$O lines.
The validity of this disk simplification 
will be discussed later
by comparison with spherical models (Sect.~\ref{spherical}).

 We note that `two layer slab model', or even spherically geometric two layer
model, does not describe the physics in the atmosphere of the star.
These models are used to fit the observed spectra,
and to interpret the observational results. 
Two layers are assumed to be
independent and discontinuous without any interaction,
which are unlikely in the real atmosphere.

For the energy level populations of the molecules,
local thermodynamic equilibrium (LTE) is assumed.
A turbulent velocity of 5\,km\,s$^{-1}$ is assumed for all the molecules
(see Yamamura et al.~\cite{Yamamura99b}).
 The line list of H$_2$O is taken from 
Partridge \& Schwenke (\cite{Partridge97})
and the isotope ratio from HITRAN~1996
(Rothman et al.~\cite{Rothman99}) is applied.
The line lists of OH and CO$_2$,
which are used for data-analysis,
are also taken from HITRAN.

\begin{table}
\begin{caption}
{H$_2$O bands of main isotope 
in near-infrared region (Herzberg~\cite{Herzberg45}).
H$_2$O has three vibrational levels;
symmetric stretch ($v_1$),
bending of the OH bond ($v_2$),
and asymmetric stretch ($v_3$).
The lower state 
of vibrational levels is 0,0,0 for all bands.
}
\label{table-H2O lines}
\end{caption}
\begin{center}
\begin{tabular}{lllcc} \hline
\multicolumn{2}{c}{Band centre} &
\multicolumn{3}{c}{Upper state} \\ \cline{1-2}\cline{3-5}
[cm$^{-1}$] & [$\mu$m] & $v_1$ &$v_2$ & $v_3$ \\ \hline
1594.8 & 6.270 &0&1&0 \\
3151.4 & 3.173 &0&2&0 \\
3657.1 & 2.734 &1&0&0\\
3755.8 & 2.663 &0&0&1\\\hline
\end{tabular}
\end{center}
\end{table}
\begin{table}
\begin{caption}{
The approximate H$_2$O column densities where optical depth $\tau$ becomes
unity at each wavelength.
The turbulent velocity is 5\,km\,s$^{-1}$.}
\label{table-H2O tau}
\end{caption}
\begin{tabular}{r rrrr} \hline
$T_{\mathrm {ex}}$ & $N_{\mathrm \tau_{3.8\,\mu m}=1}$ & 
$N_{\mathrm \tau_{3.5\,\mu m}=1}$ & $N_{\mathrm \tau_{3.2\,\mu m}=1}$ & 
$N_{\mathrm \tau_{2.5\,\mu m}=1}$ \\ 
 & [cm$^{-2}$] & [cm$^{-2}$] & [cm$^{-2}$] & [cm$^{-2}$] \\ \hline
2000 & $5.5\times10^{21}$ & $1.4\times10^{21}$ & $3.5\times10^{20}$ & $1.3\times10^{20}$\\
1400 & $2.4\times10^{22}$ & $3.4\times10^{21}$ & $7.6\times10^{20}$ & $1.2\times10^{20}$\\
1200 & $4.9\times10^{22}$ & $5.4\times10^{21}$ & $1.0\times10^{21}$ & $1.3\times10^{20}$\\
1000 & $1.1\times10^{23}$ & $1.0\times10^{22}$ & $1.3\times10^{21}$ & $1.5\times10^{20}$\\ 
 800 & $3.3\times10^{23}$ & $2.6\times10^{22}$ & $1.7\times10^{21}$ & $2.1\times10^{20}$\\
 600 & $1.6\times10^{24}$ & $1.1\times10^{23}$ & $2.1\times10^{21}$ & $3.7\times10^{20}$\\ 
 400 & $1.6\times10^{25}$ & $1.2\times10^{24}$ & $2.3\times10^{21}$ & $1.3\times10^{21}$\\
\hline
\end{tabular}
\end{table}
\begin{table}
\begin{caption}
{
Results derived by Yamamura et al. (\cite{Yamamura99b}).
}
\label{table-yamamura}
\end{caption}
\begin{tabular}{r cc cc cc} \hline
Name & \Thot & \Tcool &log \Nhot & log \Ncool & \Rhot & \Rcool \\ 
 & [K] & [K] & [cm$^{-2}$] & [cm$^{-2}$] & [$R_*$] & [$R_*$]\\\hline
$o$~Cet & 2000$^{\dagger}$ & 1400 & 21.0 & 20.5 & 2.0 & 2.3 \\
Z~Cas   & 2000$^{\dagger}$ & 1200 & 22.0 & 21.0 & 1.1 & 1.7 \\ \hline
\end{tabular} \\
$^{\dagger}$ assumed.
\end{table}

\subsection{The model parameters}
 We calculated more than 50\,000 spectra
covering a wide range of parameters
in order to fit the observed spectra, using the $\chi^2$ analysis.
The ranges and step sizes of the calculated parameters are summarized 
in Table~\ref{tbl-parameter-range}.
There are seven parameters.
$T$, $N$, $R$ of the `hot' and `cool' layers are given independently.
The background continuum is assumed to be a blackbody
of temperature $T_*$ of 3000\,K.
In the following section
the effects of each parameter on the spectra
are described in detail.
We emphasize that the radius of the hot layer (\Rhot)
is an essential parameter for the variation between absorption and emission
at wavelengths longer than $\sim$3.8\,$\mu$m.
The other parameters are less important
to the features in this wavelength range.
In the following figures (Figs.~\ref{Tcen2}--\ref{Nc})
the model spectra are presented
for a star at a distance of 100\,pc and with a radius $R_*$ of
$3\times10^{13}$\,cm.

\begin{table}
\begin{caption}
{
The ranges of the parameters for $\chi^2$ analysis.
}
\label{tbl-parameter-range}
\end{caption}
\begin{tabular}{l c r l} \hline
Parameter & Values & Grid & Comments \\ \hline
$T_*$ [K] & 3000 & & Fixed \\
\Thot\ [K] & 2000 & & Fixed \\
\Tcool\ [K] & 1000--1400 & 100 & \\
log \Nhot [cm$^{-2}$] & 20.00--23.00 & 0.25 & \\
log \Ncool [cm$^{-2}$] & 20.00--23.00 & 0.25 & \\
\Rhot [$R_*$] & 1.0--3.0 & 0.2 & \Rhot$<$\Rcool \\
\Rcool [$R_*$] & 1.0--3.0 & 0.2 & \\
            & 3.0--4.5 &   0.5 & \\ \hline
\end{tabular} 
\end{table}


\subsubsection{The spectrum of the central star as background source}
\begin{figure}[!ht]
\resizebox{\hsize}{!}{\includegraphics*[30,370][555,720]{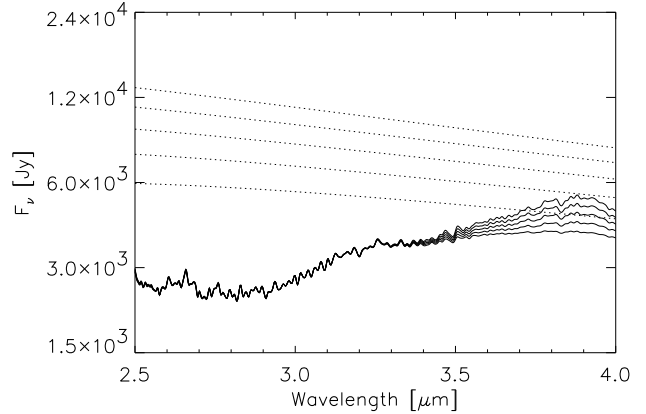}}
\caption{
The effects of $T_*$ on the spectra are shown.
The thick lines indicate the synthetic spectra with a two-layer `slab' model,
and the dotted lines indicate the background continuum level for each $T_*$.
From top to bottom,
$T_*$=3000, 2800, 2600, 2400, 2200\,K.
Other parameters are fixed as
\Thot=2000\,K,
\Tcool=1200\,K,
\Rhot=1.0\,$R_*$,
\Rcool=1.4\,$R_*$,
\Nhot=$1\times10^{22}$\,cm$^{-2}$, and
\Ncool=$1\times10^{21}$\,cm$^{-2}$.
Due to the large optical depth of H$_2$O molecules,
the difference in background appears
only above $\sim$3.5\,$\mu$m.
}
   \label{Tcen2}
\end{figure}

Generally, the column densities of H$_2$O layers
in Mira variables (e.g. Yamamura et al.~\cite{Yamamura99b})
give optically thick layers
at most wavelengths under consideration.
In Fig.~\ref{Tcen2}, 
we show calculated H$_2$O spectra with various temperatures ($T_*$)
of the blackbody as a background spectrum of H$_2$O,
and we examine the resultant dependence of H$_2$O bands
on the background source.
This figure demonstrates that
the effects of the background source only appear
at wavelengths longer than $\sim$3.5\,$\mu$m
where the optical depth of H$_2$O becomes small.
In the present analysis, we assume
that the background spectrum is a blackbody with
temperature $T_*$=3000\,K,
since it is difficult to determine this parameter.

\subsubsection{The radius of the hot layer} \label{sec-radii}

\begin{figure}[!ht]
\resizebox{\hsize}{!}{\includegraphics*[30,370][555,720]{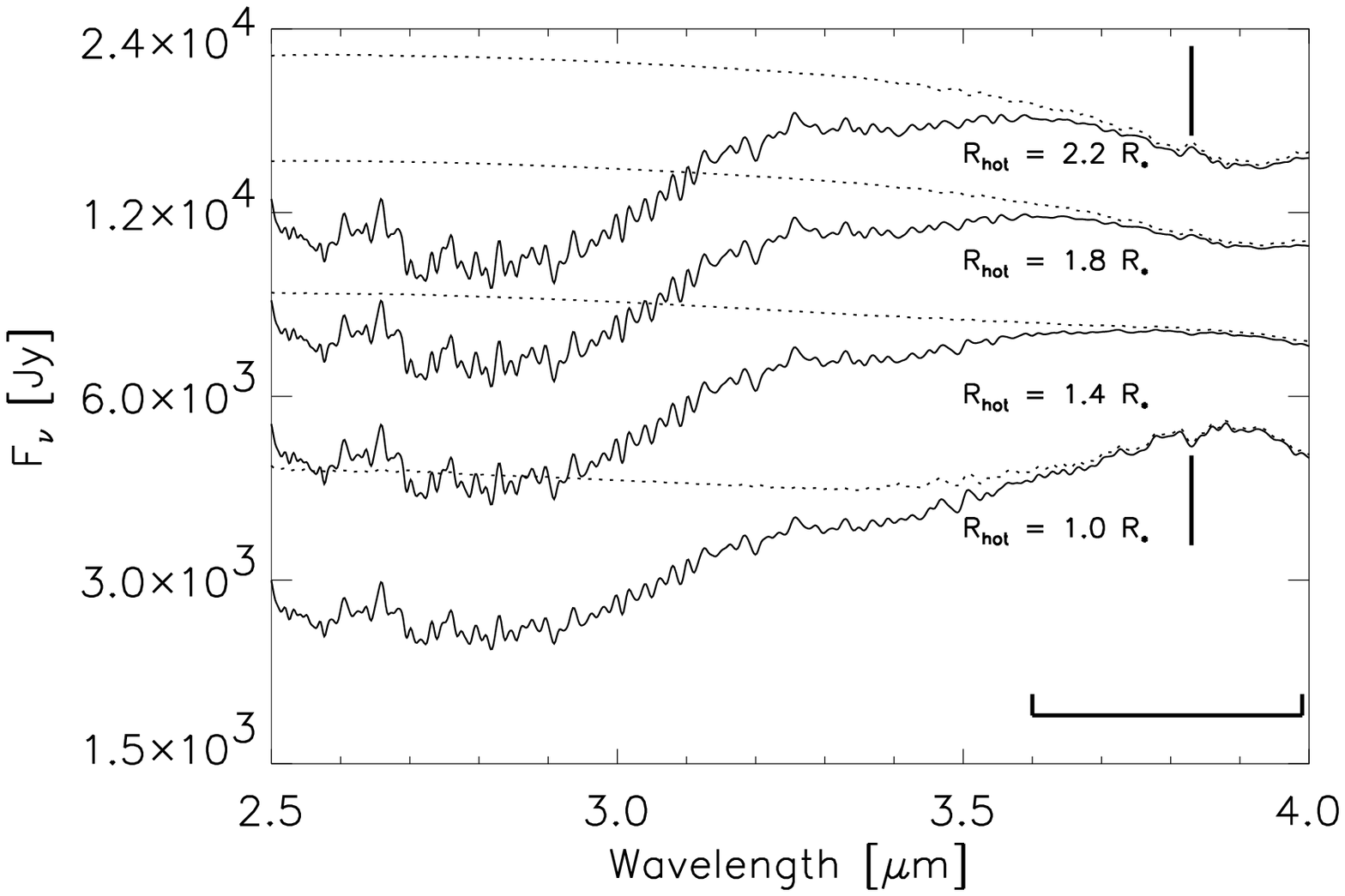}}
\begin{center}
\resizebox{0.65\hsize}{!}{\includegraphics*[10,85][290,410]{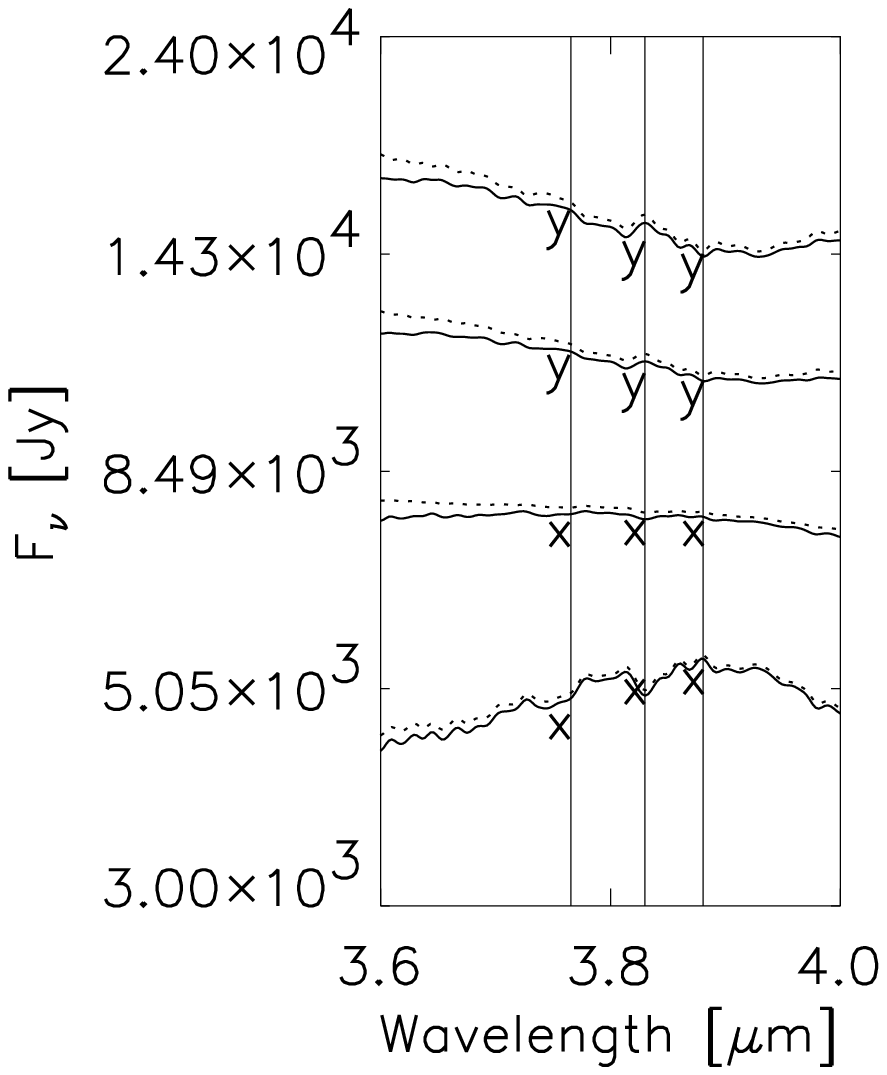}}
\end{center}
\caption{
{\it Upper:} The radius of the `hot' layer (\Rhot) 
with an excitation temperature (\Thot) of 2000\,K
influences the features longer than $\sim$3.5\,$\mu$m.
The solid lines are by two-layer model and the dotted lines
are only by the hot layer.
The effects are most prominently seen in the 3.83\,$\mu$m feature,
which is either seen in emission or in absorption depending on \Rhot.
In this case, features longer than $\sim$3.5\,$\mu$m change,
and become emission when \Rhot\ is larger than $\approx$1.6\,$R_*$.
Other parameters are
$T_*$=3000\,K,
\Thot=2000\,K,
\Tcool=1200\,K,
\Nhot=$1\times10^{22}$\,cm$^{-2}$,
\Ncool=$1\times10^{21}$\,cm$^{-2}$.
\Rcool\ is given as $\Rcool^2=\Rhot^2+1^2$\,[$R_*^2$].
{\it Lower:} The spectra of 3.8\,$\mu$m region are enlarged.
Conspicuous three features are marked.
Note that a feature at 3.88\,$\mu$m shows `emission like' feature
when spectra are in absorption (see text).
}
   \label{Rh}
\end{figure}

In Fig.~\ref{Rh}, the influence of \Rhot\ is shown. 
The features in the region longer than $\sim$3.5\,$\mu$m vary with \Rhot.
The spectrum shorter than $\sim$3.5\,$\mu$m is dominated by
the absorption by the overlaid `cool' layer
and is insensitive to the parameters of the hot layer.
When \Rhot\ is $\approx$1.6\,$R_*$ with \Thot=2000\,K,
the emission from the outer part of the layer ($r>R_*$) 
fills up the absorption
originating in the inner part of the layer ($r<R_*$) and 
the spectrum becomes almost flat
at wavelengths longer than 3.5\,$\mu$m.
Emission features appear when \Rhot\ becomes larger than $\approx$1.6\,$R_*$.
The effects are most prominently seen in the 3.83\,$\mu$m feature
in the top panel of Fig.~\ref{Rh}.
Also the slope in the 3.8\,$\mu$m region
depends on \Rhot.
The 3.8\,$\mu$m region is enlarged in the bottom panel of Fig.~\ref{Rh}.
In this plot, not only the 3.83\,$\mu$m feature but also
other two features at 3.77\,$\mu$m and 3.88\,$\mu$m are marked.
These three marked features are in emission,
when the features arising from the outer disk of \Rhot$>r>R_*$
(region {\it B} in Fig.~\ref{config})
exceed the absorption arising from the inner disk of $r<R_*$
(region {\it A} in Fig.~\ref{config}); this is marked
as `y' in the figures.
As the features arising from $r<R_*$ are dominant,
the features are inverted and we indicate the features as `x'.
These three features are not formed by a single H$_2$O line
but by a combination of a large number of overlapping H$_2$O lines.
The definition of the emission and absorption features
is opposite for the 3.88\,$\mu$m feature,
as this feature is formed by slightly smaller optical depth
compared to the neighboring regions.

\subsubsection{The excitation temperature}
The observed spectrum of $o$\,Cet suggests that \Thot\ must be above $\sim$1600\,K;
below $\sim$1600\,K, the flux level of the synthesized spectra
shortward of 2.5\,$\mu$m
is lower than the flux levels of the observed spectra.
Because the spectrum of the optically thick hot H$_2$O layer
behaves like a blackbody with temperature of \Thot,
the flux level around 2.5\,$\mu$m decreases considerably
below 1600\,K.
Thermal equilibrium chemistry shows that H$_2$O molecules are 
abundant below $\sim$2200\,K in the atmosphere of red giants
and that the H$_2$O abundance decreases significantly above this temperature
(Tsuji \cite{Tsuji64}). 
The hot layer is expected to have an excitation temperature 
in the range of $1600 < \Thot < 2200$\,K.
With the resolution of 300 it is difficult to determine \Thot.
Thus, we fix \Thot\ as 2000\,K.

\begin{figure}[!ht]
\resizebox{\hsize}{!}{\includegraphics*[30,370][555,720]{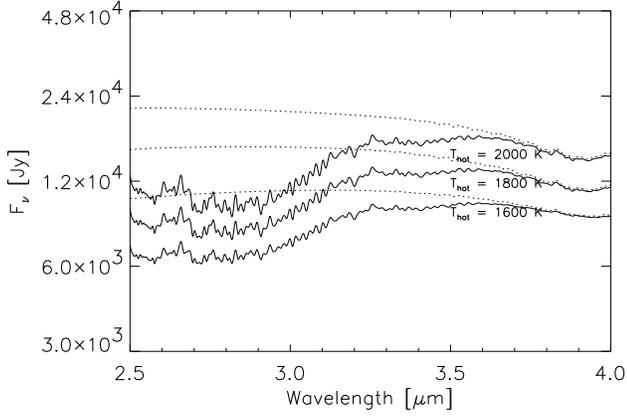}}
\caption{
The effects of \Thot\ for the case where the hot H$_2$O is seen in emission.
From top to bottom $T_{hot}$=2000, 1800, 1600\,K.
Full lines show the spectra with two-layers,
and dotted lines show the spectra
only with the hot layer as a background of the `cool' layer.
Other parameters are
$T_*$=3000\,K,
\Tcool=1200\,K,
\Rhot=2.2\,$R_*$,
\Rcool=2.4\,$R_*$,
\Nhot=$1\times10^{22}$\,cm$^{-2}$,
\Ncool=$1\times10^{21}$\,cm$^{-2}$.
}
   \label{Th}
\end{figure}

When the hot layer produces emission features,
the emergent H$_2$O features always appear similar (Fig.~\ref{Th}). 
We calculated model spectra with \Thot\ of
1800\,K and 2000\,K and confirmed
that the other parameters were only little affected.
\Rhot\ becomes systematically larger by 0.2\,$R_*$ at most
(1 parameter step of \Rhot) at \Thot=1800\,K compared to \Thot=2000\,K.

\subsubsection{The column density}\label{sec-column-density}
\begin{figure}[!ht]
\resizebox{\hsize}{!}{\includegraphics*[30,370][555,720]{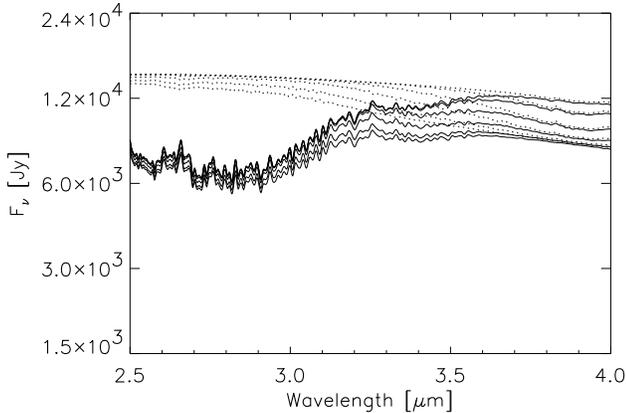}}
\caption{
The effects of \Nhot\ on the spectra
when hot H$_2$O is seen in emission.
The solid lines are for two-layer model and the dotted lines
are only for the hot layer.
From top to bottom $N_H$=$3\times10^{22}, 1\times10^{22}, 3\times10^{21},
1\times10^{21}, 3\times10^{20}, 1\times10^{20}$\,cm$^{-2}$.
Other parameters are
$T_*$=3000\,K,
\Thot=2000\,K,
\Tcool=1200\,K,
\Rhot=1.8\,$R_*$,
\Rcool=2.0\,$R_*$,
\Ncool=$1\times10^{21}$\,cm$^{-2}$.
}
   \label{Nh}
\end{figure}

Fig.~\ref{Nh} shows that \Nhot\ changes the spectra
moderately.
The noticeable features around 3.8\,$\mu$m
(Sect.~\ref{sec-radii}) are seen
when \Nhot\ is in the range of
$1\times10^{21}$--$3\times10^{22}$\,cm$^{-2}$.
Below or above this range these feature are not detectable
at the current wavelength resolution.

\begin{figure}[!ht]
\resizebox{\hsize}{!}{\includegraphics*[30,370][555,720]{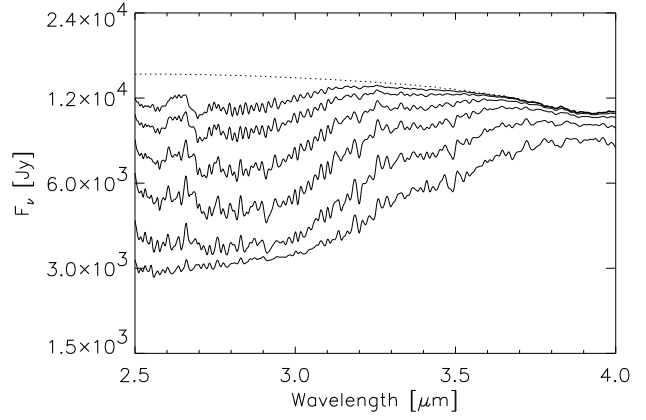}}
\caption{
The influence of \Ncool\ on the spectrum.
The solid lines are for two-layer model and the dotted line
is for the hot layer as the background of the cool layer.
From top to bottom,
\Ncool=$1\times10^{20}, 3\times10^{20}, 1\times10^{21},
3\times10^{21}, 1\times10^{22}, 3\times10^{22}$\,cm$^{-2}$.
Other parameters are
$T_*$=3000\,K,
\Thot=2000\,K,
\Tcool=1200\,K,
\Rhot=1.8\,$R_*$,
\Rcool=2.0\,$R_*$,
\Nhot=$1\times10^{22}$\,cm$^{-2}$.
}
   \label{Nc}
\end{figure}

The wavelengths shorter than 2.7\,$\mu$m are dominated by the lines 
in the R-branches, while the lines in P-branches start at longer wavelengths.
The gap between the two branches is seen as an `emission-like feature'
at 2.7\,$\mu$m.
The column density of the cool layer (\Ncool) determines
the feature around 2.7\,$\mu$m (Fig.~\ref{Nc}).
T~Cep and R~Aql, with column densities
in the order of $10^{20}$--$10^{21}$\,cm$^{-2}$
show the feature.
R~Cas and Z~Cyg, whose features around 2.7\,$\mu$m
cannot be clearly seen, have higher column densities
than this range.


\subsection{Other molecules in this wavelength range}
\subsubsection{OH}

 There are sharp OH bands visible 
in some spectra of R~Aql and T~Cep,
especially around visual maximum.
The Q-branch lines at 2.8~$\mu$m do not appear strongly
in the observed spectra (Fig.\,\ref{obs}).
This is because H$_2$O with its large optical depth
masks the OH Q-branch lines in the 2.8\,$\mu$m region.

In the current analysis, we include OH molecules only in the cool layer,
and we assume that the excitation temperature and 
radius for OH molecules are the same as that of cool H$_2$O layer
so as to reduce the number of the parameters.
We confirm that the H$_2$O parameters are not changed
by more than the typical uncertainties by including OH molecules.

 OH molecules can persist up to 4000\,K,
which is higher temperature than for H$_2$O
(Tsuji~\cite{Tsuji64}).
The assumption of the
OH excitation temperature in our model is not ideal.
The OH lines with higher excitation temperature than the hot H$_2$O layer
are not seen in the model spectra due to large opacity of H$_2$O.
We ignore OH with higher excitation temperature.

\subsubsection{CO$_2$}
\begin{figure}[!ht]
\begin{center}
\resizebox{6cm}{!}{\includegraphics*[90,170][515,615]{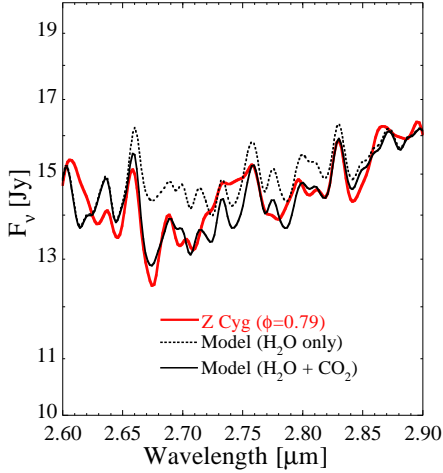}}
\end{center}
\caption{
CO$_2$ features around 2.7\,$\mu$m.
An example is the spectrum in Z~Cyg-2 ($\phi$=0.79).
We put one extra CO$_2$ layer
over the cool H$_2$O layer in the `slab' model.
The parameters of synthetic CO$_2$ spectrum are
estimated to be excitation temperature of 800\,K, a column density of 
$1\times10^{19}$\,cm$^{-2}$, and a radius of 4.0\,$R_*$.
The parameters of H$_2$O are listed in Table~\ref{tbl-results}.
}
   \label{zcyg2_co2}
\end{figure}

 $^{12}$C$^{16}$O$_2$ absorption of
the $10^01$--$00^00$ and $02^01$--$00^00$ vibration bands
(Herzberg~\cite{Herzberg45})
is seen at 2.7\,$\mu$m.
The CO$_2$ spectrum of Z~Cyg ($\phi=0.79$) is indicated
in Fig.~\ref{zcyg2_co2}.
CO$_2$ molecules are superposed over
the two H$_2$O layers.
The feature at 2.67\,$\mu$m is reproduced well by CO$_2$.
The parameters of the synthetic CO$_2$ spectrum in Fig.~\ref{zcyg2_co2} are
an excitation temperature of 800\,K, a column density of 
$1\times10^{19}$\,cm$^{-2}$, and a radius of 4.0\,$R_*$.
These parameters are estimates only.
We mask the wavelength region of 2.65--2.75\,$\mu$m
for the $\chi^2$ analysis of the H$_2$O fitting and do not discuss 
the CO$_2$ molecules in this paper.

\subsubsection{SiO}
 SiO $\Delta v$=2 bands at 4.1\,$\mu$m region are 
present in the SWS spectra.
The $\Delta v$=3 vibration levels around 3\,$\mu$m region are not clear.
We mask the wavelengths longer than 3.95\,$\mu$m 
in the $\chi^2$ analysis in order to avoid
the influence of the SiO molecules.

\subsection{Spherical effects}\label{spherical}
 We adopted disk-shape plane-parallel models
to calculate a number
of spectra with various parameter sets for the $\chi^2$ analysis.
We examine the parameters derived from the `slab' model.
In order to validate this simplification,
spectra of spherical models are compared 
with `slab' model spectra using the same $\chi^2$ analysis.
In the spherical geometry, two H$_2$O layers with uniform temperatures
are adopted.
The density gradient in each layer is assumed to vary as $r^{-2}$.
An example is indicated in Table~\ref{slab-spherical}.
The parameters are consistent between these two models.
The results derived from the simple disk model 
can provide fairly good estimates of the structure of the outer layers.
This is because the optical depth of the H$_2$O layer is so high 
(optical depth at 3.5\,$\mu$m becomes about unity with a column density of
$1\times 10^{21}$\,cm$^{-2}$ and an excitation temperature of 2000\,K; Table~\ref{table-H2O tau}), 
that only the surface of each sphere is important for the synthetic spectra.
The radius where the optical depth becomes thick in a spherical
model is represented as the radius of the `slab' model.
If we compare the spectra with these two geometric models in an optically
thin case (say column density of $10^{18}$--$10^{19}$\,cm$^{-2}$), 
we find larger discrepancies.

\begin{table}
\caption{
Validation of a `slab' model as a simplification of spherical model.
A spectrum of spherical model is compared 
with `slab' model using the $\chi^2$ analysis.
\Rin\ and \Rout\ are the inner boundary and the outer boundary
of the spherical model, respectively.
\Thot\ is 2000\,K for both models.
}\label{slab-spherical}
\tiny
\begin{tabular}{lcc cc cc cc}\hline
 &\Tcool&\multicolumn{2}{c}{\Rhot}
 &\multicolumn{2}{c}{\Rcool}
 &\Nhot	&\Ncool	\\
 & [K] & \multicolumn{2}{c}{[$R_*$]} & \multicolumn{2}{c}{[$R_*$]} & [cm$^{-2}$] & [cm$^{-2}$] \\

		&	&\Rin	&\Rout		&\Rin	&\Rout		&			&		\\ \hline
Spherical&1400		&1.0	&2.2				&2.2	&3.0				&$3\times 10^{21}$	&$3\times 10^{20}$	\\
`Slab'	&1400	&\multicolumn{2}{c}{2.0}	&\multicolumn{2}{c}{2.8}			&$3\times10^{21}$	&$3\times10^{20}$	\\ \hline

\end{tabular}
\end{table}

\subsection{Fitting process} \label{section-fit}
 The best fit for each observed spectrum is derived
by the $\chi^2$ analysis.
The spectral resolution is 300--500 in these wavelengths.
All the spectra are convolved by a Gaussian with
the wavelength resolution of $\lambda/\Delta\lambda=300$.
$\chi_m^2$ is the reduced $\chi^2$, and divided by the degrees
of freedom $m$. 
The best parameter sets are found by minimizing $\chi_m^2$.
The wavelength range 2.5--3.95\,$\mu$m
is used except for the 2.65--2.75\,$\mu$m region
which is strongly affected by CO$_2$ bands.

The SWS scans the same wavelength region twice.
The difference between two scans is a good indication of the observational uncertainty.
Two scans are normally consistent
within 0.6\,\% of the flux over 2.5--4.0\,$\mu$m.
In some observations AOT bands 1D and 1E (3.5--4.0\,$\mu$m) show
a significant difference between two scans. 
The difference is as large as 10\,\% of the flux in the worst case (R~Aql-1;
we discard this observation from the later discussions).
We added this difference in flux and 1\,\% of the flux, 
which represents the scatter of the data around the re-gridded data,
to an observational error, $\sigma$.
The latter part is necessary to
avoid $\chi^2$ diverging to the infinity.
The error of each parameter is obtained by calculating grid points
within 3 $\sigma$ level from the parameter of the best $\chi^2$,
when other parameters are fixed.
The error of our fit is not a continuous function but rather a grid scale.

\section{Results}

\subsection{The best-fit spectra and their parameters}
\begin{figure*}[!ht]
\begin{center}
\resizebox{\hsize}{!}{\includegraphics*[30,360][470,805]{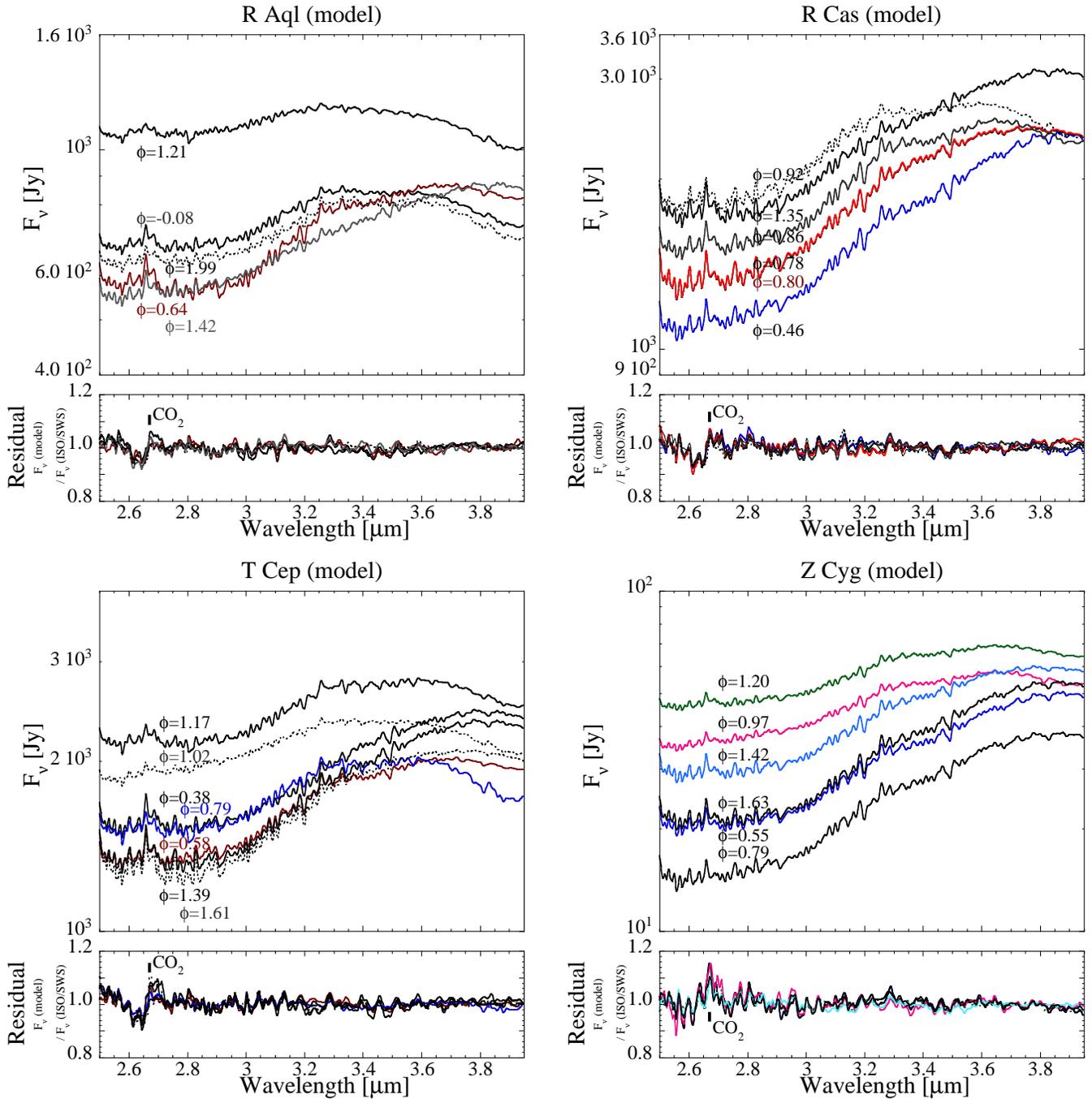}}
\end{center}
\caption{
 The fitted results and the residuals.
Two spectra at $\phi$=0.78 (on the same day)
and one spectra at $\phi$=0.80 of R~Cas
are fitted with the same parameters,
and the synthetic spectra are identical.
}
   \label{model}
\end{figure*}

\begin{figure}[!ht]
\begin{center}
\resizebox{\hsize}{!}{\includegraphics*[160,100][410,570]{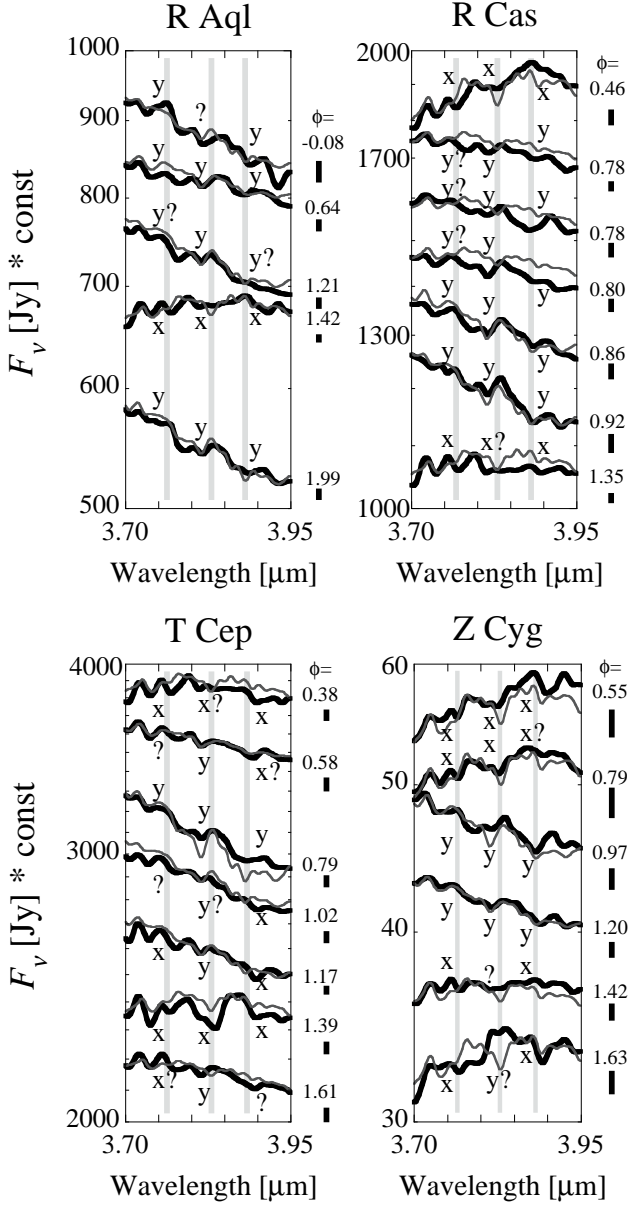}}
\end{center}
\caption{
The spectra in the 3.8\,$\mu$m region are enlarged;
the ISO/SWS spectra (bold lines),
and the model spectra (thin line).
The three conspicuous features (see Sect.~\ref{sec-radii}; Fig.~\ref{Rh}),
which shows time variation, are marked.
The mark, `x' shows the absorption phase
and `y' shows the emission phase 
of ISO/SWS spectra.
The definition of the emission and absorption feature at 3.88\,$\mu$m
is opposite to their aspects.
Around minimum the features are observed in absorption,
while around maximum they are seen in emission.
The error bars below the phase
indicate $\sigma$ (Sect.~\ref{section-fit})
at 3.83\,$\mu$m.
Slight wavelength shift of the feature peak is seen,
especially in Z~Cyg.
These shifts seem to be observational errors.
}
   \label{spec_3.8}
\end{figure}

The best fit synthetic spectra corresponding to the observed spectra
(Fig.~\ref{obs}) are shown in Fig.~\ref{model}.
Our simple `slab' model reproduces
the global shape of the observed spectra.
Residuals between 
the observed spectra and the fitted spectra
are shown in Fig.~\ref{model}.
The fit is as good as a few \%
in the most of the investigated spectral range
and about 10\,\% at the worst part.
The 3.8\,$\mu$m region of the observed spectra and the model spectra
is enlarged in Fig.~\ref{spec_3.8}.
The variation with phase in the three conspicuous features
is seen in the observed spectra, and these features are reproduced
by the model except at $\phi$=0.78 and 0.80 of R~Cas (R~Cas, R~Cas-2, R~Cas-3).
The small features of the observed spectra are generally reproduced
in respect to the trend of emission or absorption.
However,
the detailed shapes of the small
features (e.g. height or width)
are sometimes not fitted well.
This implies that two layer slab model is too simple approximation
and the atmospheric structure is more complicated in Mira variables.

The parameters of the best fit spectra
are listed in Table~\ref{tbl-results}.
All four stars show the common tendency
that the radius of the hot layer (\Rhot) is about 1.0\,$R_*$
around minimum and about 2.0\,$R_*$ around maximum (Fig.~\ref{rh_phase}),
confirming the suggestion in Yamamura et al. (\cite{Yamamura99b}).
The cool layer is located at about 2.5--4\,$R_*$.
The column density of the cool layer (\Ncool)
varies from object to object rather than showing a trend with the phase
and $\log \Ncool$ ranges from 20.75\,cm$^{-2}$ to 22.25\,cm$^{-2}$.

The column density of the hot layer (\Nhot) has a large uncertainty
as discussed in Sect.~\ref{sec-radii}.
From the equivalent width, we could give
a lower limit when the  features are clearly seen as emission.
At the phases with the strongest emission features at 3.83\,$\mu$m
in each star
(R~Aql-5 at $\phi$=1.99; R~Cas-5 at $\phi$=0.92;
T~Cep-3 at $\phi$=0.79; Z~Cyg-3 at $\phi$=0.97),
the column density of the hot layer should be
higher than $\approx$$10^{21}$\,cm$^{-2}$,
otherwise the features would not have been detected.


 There is an uncertainty remaining
in the fits of individual spectra. 
The spectra of R~Cas, R~Cas-2, R~Cas-3,
which were taken within 2 weeks,
are fitted by the same parameter sets.
These three synthetic spectra based on $\chi^2$ analysis 
show absorption features in the 3.8\,$\mu$m region.
However, all of them show emission features in the observed spectra
(Fig.~\ref{spec_3.8}). Thus, the fitted parameters for these
three spectra are not reliable.
Also the features in the 3.8\,$\mu$m region in Z~Cyg-6 are not reproduced
with the best parameters. 
However, the data of Z~Cyg-6 are not affected by the instrumental error;
the features in the 3.8\,$\mu$m region in Z~Cyg-6
seem to be real.
Thus, the parameters of Z~Cyg-6 are doubtful.

 In some phases,
the features in the 3.8\,$\mu$m region do not clearly 
appear in the observed spectra, for example, in T~Cep-2 ($\phi$=0.58).
There are at least three solutions:
(1) The column density of the hot layer is extremely high,
and the H$_2$O lines are completely saturated.
(2) The radius of the hot layer just falls at $\approx$1.6\,$R_*$ where
the contributions of emission fills up the absorption features.
(3) The hot layer disappears, i.e. the column density of 
the hot layer is so low that the features are undetectable.
We cannot distinguish the three cases from the observed spectra.
Although the $\chi^2$ analysis suggests that case (1)
is the most likely for most of such spectra,
cases (2) and (3) cannot not be ruled out.

\subsection{The 3.8\,$\mu$m region}

The most interesting parameter in this study is
the radius of the hot layer, \Rhot.
The observed spectra in
Fig.~\ref{spec_3.8} indicate
that the 3.8\,$\mu$m features show emission around maximum
and absorption around minimum.
This suggests that \Rhot\ is larger than $\approx$1.6\,$R_*$ around
maximum and smaller than that around minimum.
The features in emission phase are not always exactly 
the inverse shape of the absorption features,
possibly because of the complicated structure of the atmosphere in Mira variables.

\begin{figure}[!ht]
\resizebox{\hsize}{!}{\includegraphics*[50,220][520,590]{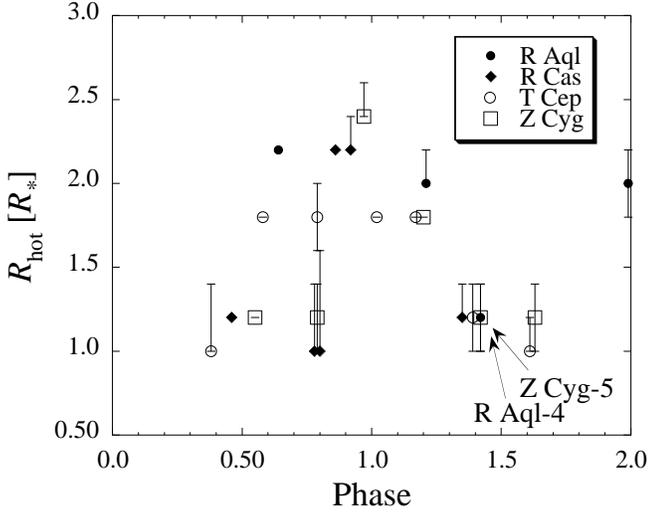}}
\caption{
The time variation in the radius of the hot layer (\Rhot).
When \Rhot\ is larger than 1.6\,$R_*$, the spectrum
around 3.8\,$\mu$m is in emission,
while \Rhot\ smaller than 1.6\,$R_*$ indicate
absorption features.
All four stars indicate about 2.0\,$R_*$ around maximum
and about 1.0\,$R_*$ around minimum.
}
   \label{rh_phase}
\end{figure}

 The variation in \Rhot\ derived from the model fitting is
plotted in Fig.~\ref{rh_phase} against the optical phase.
In all four cases, \Rhot\ reaches about 2\,$R_*$ around maximum
and becomes $\sim$1\,$R_*$ around minimum.

\begin{figure}[!ht]
\begin{center}
\resizebox{\hsize}{!}{\includegraphics*[35,165][510,600]{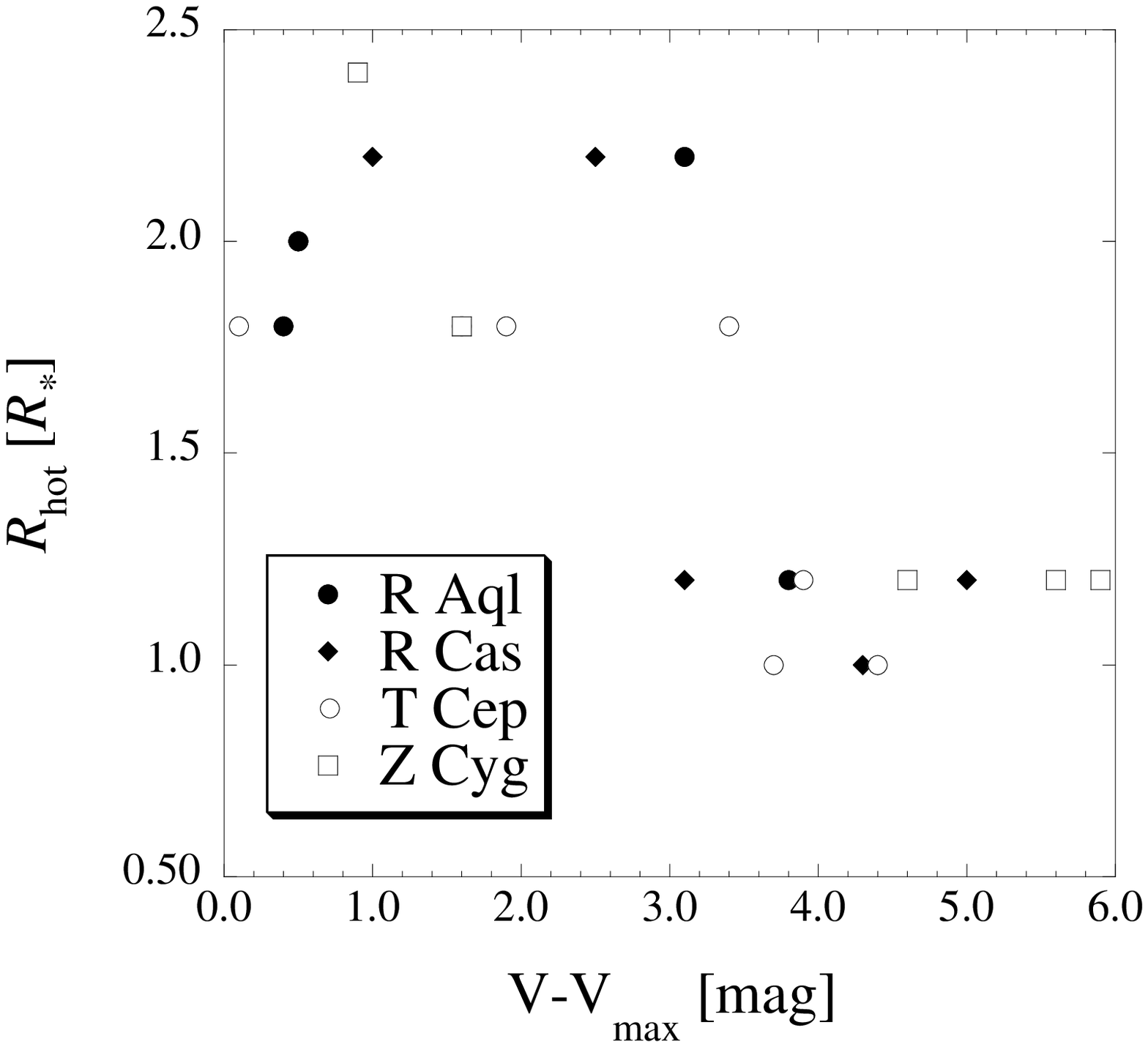}}
\end{center}
\caption{
\Rhot\ as a function of visual magnitude at each epoch
subtracted from maximum magnitude.
There is a separation;
when $\mathrm{V}-\Vmax$ is larger than 3.0\,mag
\Rhot\ is $\sim$1.0\,$R_*$, and smaller than 3.0\,mag
it is $\sim$2.0\,$R_*$.
}
   \label{rh_v}
\end{figure}

 There are slight differences among the stars.
Z~Cyg still shows absorption features at phase 0.79,
while the other three stars already show weak emission.
This trend appears to be related to 
the characteristics of the light curve (Fig.~\ref{aavso}).
The light curve of Z~Cyg shows a steep increase and a gradual decrease
in the V-band.
The visual magnitude of Z~Cyg at $\phi$=0.79 is near the minimum,
while in other stars the visual magnitudes
already start increasing at this phase.
Fig.~\ref{rh_v} shows \Rhot\ as a function of visual magnitude
at each epoch,
after subtraction of the maximum magnitude.
A separation is found at $\mathrm{V}-\Vmax$ of 3.0\,mag;
below that \Rhot\ is $\sim$2\,$R_*$ and above that
\Rhot\ is $1.0\,R_*$, indicating the variation in \Rhot\
is connected with variability.
One of the classification criteria of Mira variables is that 
visual amplitude is larger than 2.5\,mag.
Fig.~\ref{rh_v} implies that 
the emission features might be related 
with the large-amplitude pulsating variables.

\begin{table*}
\begin{caption}
{The best fit parameters, and the estimated uncertainties.
OH is included in some fits
with a fixed column density of $1\times10^{21}$\,cm$^{-2}$.
The numbers of errors are not given when 
the errors are smaller than 1 step of the model parameters.}
\label{tbl-results}
\end{caption}
\begin{tabular}{lr l ll ll l cc} \hline

 Name & Phase & \Tcool & \Rhot & \Rcool & log \Nhot & log \Ncool & log $N_{\mathrm{OH}}$ & $\chi_m^2$  \\
 & & [K] & [$R_*$] & [$R_*$] & [cm$^{-2}$] & [cm$^{-2}$] & [cm$^{-2}$] &  \\ \hline \vspace{0.1cm}
R~Aql-1 & $-0.08$ & 1400 & $1.8_{-0.2}^{+0.2}$ & $2.4_{-0.2}^{+0.2}$ & 
 21.75$_{-0.50}^{+0.50}$ & $21.25_{-0.50}^{+1.00}$ &      & 1.1  \\ \vspace{0.1cm}
R~Aql-2 & 0.64 & 1400 & 2.2 & $2.6^{+0.2}$  & 
 22.50$_{-0.25}^{+0.50}$ & $21.25_{-0.25}^{+0.25}$ &      & 2.5 \\ \vspace{0.1cm}
R~Aql-3 & 1.21 & 1400 & $2.0^{+0.2}$ & $3.0_{-0.2}$  & 
 22.25$_{-0.25}^{+0.75}$ & $20.75_{-0.50}^{+0.50}$ & 21 & 2.2 \\ \vspace{0.1cm}
R~Aql-4 & 1.42 & 1200 & $1.2_{-0.2}^{+0.2}$ & 2.4  & 
 21.50$_{-0.75}^{+0.25}$ & $21.50_{-0.50}^{+1.25}$ &      & 2.4 \\ \vspace{0.1cm}
R~Aql-5 & 1.99 & 1400 & $2.0_{-0.2}^{+0.2}$ & $2.6^{+0.2}$  & 
 21.75$_{-0.25}^{+0.25}$ & $21.50_{-0.25}^{+0.75}$ & 21 & 1.5 \\ \vspace{0.1cm}
R~Cas-1 & 0.46 & 1100 & 1.2 & 2.4  & 
 21.75$_{-0.50}^{+0.25}$ & $22.00_{-0.25}^{+0.75}$ &      & 2.7 \\ \vspace{0.1cm}
R~Cas  & 0.78 &  1000 & $1.0^{+0.4}$ & 3.5  & 
 20.75$_{-0.75}^{+0.25}$ & $22.00_{-0.25}^{+0.25}$ &      & 2.9 \\ \vspace{0.1cm}
R~Cas-2 & 0.78 & 1000 & $1.0^{+0.4}$ & 3.5  & 
 20.75$_{-0.75}^{+0.50}$ & $22.00_{-0.25}^{+0.25}$ &      & 3.1 \\ \vspace{0.1cm}
R~Cas-3 & 0.80 & 1000 & $1.0^{+0.6}$ & 3.5  & 
 20.75$_{-0.75}^{+0.25}$ & $22.00_{-0.25}^{+0.25}$ &      & 3.2 \\ \vspace{0.1cm}
R~Cas-4 & 0.86 & 1200 & 2.2 & 3.5  & 
 22.25$_{-0.25}^{+0.50}$ & $21.75_{-0.25}^{+0.50}$ &      & 2.5 \\ \vspace{0.1cm}
R~Cas-5 & 0.92 & 1100 & $2.2^{+0.2}$ & 4.5  & 
 22.25$_{-0.25}^{+0.75}$ & $21.50_{-0.25}^{+0.25}$ &      & 2.3 \\ \vspace{0.1cm}
R~Cas-6 & 1.35 & 1100 & $1.2^{+0.2}$ & 2.6  & 
 21.50$_{-0.75}^{+0.25}$ & $21.50_{-0.25}^{+0.25}$ &      & 3.1 \\ \vspace{0.1cm}
T~Cep-1 & 0.38 & 1200 & $1.0^{+0.4}$ & 2.4  & 
 21.00$_{-0.50}^{+0.25}$ & $21.50_{-0.50}^{+0.50}$ &      & 4.1 \\ \vspace{0.1cm}
T~Cep-2 & 0.58 & 1300 & 1.8 & 2.6  & 
 22.50$_{-0.50}^{+0.50}$ & $21.50_{-0.25}^{+0.50}$ &      & 2.2 \\ \vspace{0.1cm}
T~Cep-3 & 0.79 & 1300 & $1.8_{-0.2}^{+0.2}$ & 2.6  & 
 21.50$_{-0.50}^{+0.25}$ & $21.50_{-0.25}^{+0.25}$ & 21 & 2.3 \\ \vspace{0.1cm}
T~Cep-4 & 1.02 & 1300 & 1.8 & 3.0  & 
 22.00$_{-0.25}^{+0.50}$ & $21.50_{-0.50}^{+0.75}$ & 21 & 2.5 \\ \vspace{0.1cm}
T~Cep-5 & 1.17 & 1400 & 1.8 & 2.4  & 
 22.00$_{-0.25}^{+0.50}$ & $21.25_{-0.25}^{+0.50}$ & 21 & 3.4 \\ \vspace{0.1cm}
T~Cep-6 & 1.39 & 1200 & $1.2_{-0.2}^{+0.2}$ & 2.2  & 
 21.25$_{-1.00}^{+0.25}$ & $21.50_{-0.25}^{+0.50}$ &      & 3.7 \\ \vspace{0.1cm}
T~Cep-7 & 1.61 & 1000 & $1.0^{+0.2}$ & 3.5  & 
 20.75$_{-0.25}^{+0.25}$ & $21.50_{-0.25}^{+0.25}$ &      & 2.7 \\ \vspace{0.1cm}
Z~Cyg-1 & 0.55 & 1100 & 1.2 & 2.2  & 
 21.75$_{-0.50}^{+0.25}$ & $22.00_{-0.25}^{+0.75}$ &      & 2.7 \\ \vspace{0.1cm}
Z~Cyg-2 & 0.79 & 1000 & $1.2_{-0.2}^{+0.2}$ & $2.8_{-0.2}$  & 
 21.50$_{-1.00}^{+0.25}$ & $22.25_{-0.25}^{+0.50}$ &      & 3.0 \\ \vspace{0.1cm}
Z~Cyg-3 & 0.97 & 1200 & $2.4^{+0.2}$ & 4.0  & 
 22.25$_{-0.25}^{+0.75}$ & $22.00_{-0.25}^{+0.50}$ &      & 1.5 \\ \vspace{0.1cm}
Z~Cyg-4 & 1.20 & 1300 & 1.8 & 2.6  & 
 22.00$_{-0.50}^{+1.00}$ & $21.75_{-0.25}^{+0.50}$ &      & 1.2 \\ \vspace{0.1cm}
Z~Cyg-5 & 1.42 & 1100 & $1.2_{-0.2}^{+0.2}$ & 2.6  & 
 21.00$_{-1.00}^{+0.50}$ & $22.00_{-0.25}^{+0.50}$ &      & 2.3 \\ \vspace{0.1cm}
Z~Cyg-6 & 1.63 & 1100 & $1.2_{-0.2}^{+0.2}$ & 2.2  & 
 21.50$_{-0.75}^{+0.50}$ & $22.00_{-0.25}^{+0.50}$ &      & 3.0 \\\hline
\end{tabular} 
\end{table*}


\section{Discussion}
\subsection{The dynamical motion in the extended atmosphere}
The spectra of H$_2$O bands in the 3.8\,$\mu$m region are reproduced
well assuming \Thot=2000\,K.
That temperature indicates that
the hot layer traces a relatively inner part of
the extended atmosphere.
The variation in \Rhot\ 
indicates the variability of the structure in this region.

The variation in \Rhot\ can be interpreted by
the variation in the density structure in the extended atmosphere
due to pulsations.
Pulsations create local high-density shells or shocks
in the extended atmosphere.
If the gas temperature in the shock is
high enough to dissociate H$_2$O molecules,
they are re-formed behind the shock front:
in this case \Rhot\ traces the region with the highest H$_2$O density
located behind the shock front.
Duari et al. (\cite{Duari99}) indicated
that H$_2$O can be reformed in a post-shock region.
If the shock is moderate and H$_2$O molecules do not dissociate, 
the H$_2$O density increases at the shock front
and \Rhot\ traces the motion 
of the shock front.
In either case, the variation in \Rhot\ traces the motion
of shocks.

\begin{figure}[!ht]
\resizebox{\hsize}{!}{\includegraphics*[0,40][590,800]{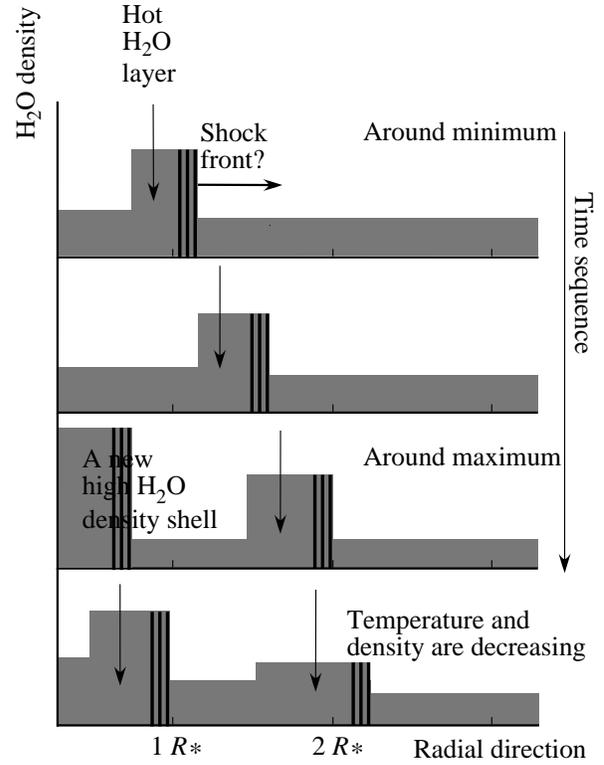}}
\caption{
A schematic view of the motion of the high H$_2$O density shell.
From top to bottom, it is in the time-sequence.
Around minimum, a high-density shell is located
on the surface of the star and H$_2$O is observed in absorption.
The high density shell is moving outwards, and located at $\approx$2\,$R_*$
around maximum.
The emission features arise from
the high density region expanded further from the `photosphere'.
After the maximum, the high density shell continues to go outwards.
As the shell is expanded, the temperature and density decreases
and its contribution to the spectrum in 3.8\,$\mu$m region
is getting lower.
}
   \label{schematic}
\end{figure}

Around minimum, a high H$_2$O density shell is created by the pulsation shock
near the stellar surface;
the area of the extended region is smaller,
and 
the contribution of the emission component originating from the extended
region is weaker than the
absorption component (Fig.~\ref{schematic}).
Toward maximum, the high H$_2$O density shell moves
away from the stellar surface.
The hot layer is extended
and emission is observed.
This high density shell continues to expand after the maximum,
but its temperature and density decrease. 
The contribution of this shell to the features in the 3.8\,$\mu$m region
gradually decreases.
At the next minimum, a new shock is formed
on the surface and produces the absorption features.

 The radius of the hot layer is measured
in units of $R_*$; the radius of the background continuum 
source (in practice, a 3000\,K blackbody).
It is uncertain how the stellar radius varies with phase.
Interferometric observations show that
the stellar radius of Mira variables is rather stable,
and varies only by less than 7\,\% at 902\,nm,
a wavelength little contaminated
with absorption
(Tuthill et al.~\cite{Tuthill95}).
The Rosseland radius derived from theoretical models
(Bessell et al.~\cite{Bessell96})
varies by about 20\,\%; the radius is smallest at maximum.
These variations in the stellar radius
are smaller than our derived variations in \Rhot/$R_*$, 
which varies by a factor of two.
Thus, the location of the hot H$_2$O layer
actually varies by as much as a factor of 1.5--2.0. 

 Hydrodynamic model calculations
indicated that H$_2$O molecules are in non-LTE
(Woitke et al.~\cite{Woitke99}).
We cannot discard non-LTE excitation for the cause of emission features.
Recently, Schweitzer et al. (\cite{Schweitzer00})
developed a method to solve non-LTE for H$_2$O molecules.
Such an effect should be considered for future work.

\subsubsection{Comparison with the structural variation in the extended
atmosphere traced by other molecules} 
 Hinkle et al. (\cite{Hinkle82}, \cite{Hinkle84})
and Lebzelter et al. (\cite{Lebzelter99})
found periodical variations
in the near-infrared CO lines in the H-band and K-band
of several Mira variables.
Hinkle et al. (\cite{Hinkle82}, \cite{Hinkle84}) found that
the CO $\Delta v=3$ lines, with excitation temperatures
of 2200--4000\,K, show a P~Cyg profile;
the blue-shifted emission feature is visible from $\phi$=0.6 to 0.8.
The CO column density is lowest around $\phi$=0.85.
These variations are clearly different from those of hot H$_2$O.
The dissociation energies of CO and H$_2$O molecules
are 11.09 and 5.11\,eV, respectively
(D\"appen~\cite{Daeppen99}).
It is reasonable to assume that the CO molecules in 
Hinkle et al. (\cite{Hinkle82}, \cite{Hinkle84})
trace even a hotter
region interior to the hot H$_2$O layers.
The authors explained the behavior of the CO lines
with the shocks caused by the pulsation.
A shock affects the CO layer around $\phi$=0.7.
The high density region caused by the shock influences
the H$_2$O layer around maximum.
It is qualitatively
reasonable that this phase lag of the expected shock penetration
appears between the CO and H$_2$O layers
because of the difference in transition energy and
binding energy of these molecules.
 
 Yamamura et al. (\cite{Yamamura99a}) found SO$_2$
in oxygen-rich Mira variables and examined the time variation.
Unlike H$_2$O,
the time variation in the SO$_2$ band does not correlate with the visual variability
but changes from emission to absorption over more than a single period.
The location of the SO$_2$ layer is estimated to be about 4--5\,$R_*$ in T~Cep.
It is implied that 
the inner region in the extended atmosphere (hot H$_2$O layer) follows 
the optical variability, but the outer region (SO$_2$ layer) does not.
It cannot be ruled out that the outer region
has a longer variability period
than the inner region
because the pulsation shocks do not directly influence the outer region.

\subsubsection{Comparison with the hydrodynamic models}

 Bessell et al. (\cite{Bessell96}) calculated atmospheres
of oxygen-rich Mira variables including hydrodynamics.
They show CO line profiles for several transitions.
The CO first-overtone (excitation level of 0.2\,eV) is 
seen in emission at $\phi$=0.8--1.0.
They explained that the CO emission was produced in the pre-
and post-shock region in the inner part of the extended atmosphere.
Winters et al. (\cite{Winters00}) calculated the CO line profile
for both $\Delta v = 1$ and $\Delta v=2$ transitions. 
CO lines, especially those of $\Delta v=1$, show a P-Cyg profile
with a strong emission component.
Although their model is calculated for carbon stars,
the general trend may be applied to oxygen-rich stars.
These theoretical models indicate that CO lines are seen in emission
due to the high density regions of CO molecules lifted by the pulsation shocks.
The appearance of H$_2$O lines can be interpreted in a similar manner.

 Bessell et al. (\cite{Bessell96}) showed
the temperature and density structures in the extended atmosphere.
A shock occurs at the photosphere and moves outwards.
Only models for the fundamental mode of pulsation
can produce shocks in the extended atmosphere.
Their models indicated that the temperature does not increase
drastically at the shock front
(nearly isothermal shock).
According to their Z-model 
(fundamental pulsation mode and luminosity of 6310\,$L_{\sun}$),
the location of the shock increased by a factor of 1.8 
from minimum to maximum.
The radius of the H$_2$O layer measured in our work
is a factor of 1.5--2.0 larger around maximum than around minimum.
Assuming that the hot H$_2$O layer traces either the location of shock itself
or post-shock region,
the approximate scales of the shock propagation are comparable with that
model.

Bessell et al. (\cite{Bessell96}) also calculated
the temperature and the density around the shock region.
The temperature and the density at the shocked region have
uncertainties according to Bessell et al. (\cite{Bessell96}).
Here, we discuss the post-shock region.
They show one example, D-model
(fundamental pulsation mode and luminosity of 3470\,$L_{\sun}$).
According to one of their D-models,
the gas temperature is 2000\,K and the mass density is 
$1\times10^{-11}$\,g\,cm$^{-3}$ at the post-shock region at maximum phase.
The gas temperature is comparable to 
the excitation temperature of our hot layer.
The H$_2$O number density becomes $5\times10^{8}$\,cm$^{-3}$ if we assume
the gas is dominated by H$_2$ and the abundance ratio of H$_2$ and 
H$_2$O is $10^{-4}$.
Our analysis indicates that
the column density of the hot layer is 
required to be higher than $\approx 1\times 10^{21}$\,cm$^{-2}$
to detect the emission features in the 3.8\,$\mu$m region.
The density calculated by Bessell et al. (\cite{Bessell96})
exceeds the column density $ 1\times 10^{21}$\,cm$^{-2}$
out to 0.1 stellar radii.
Here we assume $3\times10^{13}$\,cm for one stellar radius.
Thus, the density in Bessell et al. (\cite{Bessell96})
is consistent with our estimate.

 Another hydrodynamic model is presented
by Woitke et al. (\cite{Woitke99}). Unfortunately,
the phase is not given in their work, but
a shock with a gas temperature of 2000\,K reaches 2 stellar radii.
In their model, the H$_2$O number density at the shock front is
only $3\times10^{5}$\,cm$^{-3}$.
A shell thickness
larger than $10^{15}$\,cm (more than $\sim$30\,stellar radii) is required
to reach the H$_2$O column density of $1\times10^{21}$\,cm$^{-2}$.
The number density in Woitke et al. (\cite{Woitke99})
is by more than an order of magnitude lower than our results.
Their results may be applicable to the observations
of irregular variables and semi-regular variables by 
Tsuji et al. (\cite{Tsuji97}), and not to stronger pulsating stars
such as Mira variables.

\subsection{The optically thick H$_2$O layers 
in oxygen-rich Mira variables 
and their infrared variability}
 The near-infrared variability has been measured for Mira variables.
For example, the L'-band magnitude of R~Aqr varies
about 0.6\,mag (Le~Bertre~\cite{Lebertre93}).
In our sample, the variations are about a factor of 1.4--2.2
(0.5--0.7 in magnitudes) at L'-band
(Table~\ref{table-observation}).
As the opacity is dominated by H$_2$O in this wavelength range,
it is likely that H$_2$O molecules play an essential role in
the flux variation.

From Figs.~\ref{Tcen2}--\ref{Nc}, 
the parameters which drastically
vary the absolute flux at 3.8\,$\mu$m (the effective wavelength of the L'-band;
van der Bliek et al.~\cite{vanderBliek96}), 
are the radius of the hot layer (\Rhot; see Fig.~\ref{Rh})
and the excitation temperature of the hot layer 
(\Thot; see Fig.~\ref{Th}).
The variation in \Rhot\ from 1.0 to 2.0\,$R_*$
induces a variation in the emission by a factor of 4.0.
The difference in the flux between \Thot=1600 and 2000\,K
is a factor of two.
The flux  variation is explained by the larger radius
and lower excitation temperature of the high H$_2$O density
shell at maximum compared to minimum.

Tuthill et al. (\cite{Tuthill00})
measured the radius of a Mira variable with interferometry:
they mentioned that the radius at 3.1\,$\mu$m is larger than
those in other near-infrared bands, due to
molecular opacity.
The hot H$_2$O layer in our work
fits this opacity source.

 These optically thick H$_2$O layers are important for radiative
transfer in the extended atmosphere.
Model calculations of the extended atmosphere
with precise treatment of molecular opacity 
(e.g. H\"ofner~\cite{Hoefner99}; Helling~\cite{Helling00})
must be compared with the observations to 
obtain better understanding of the physical conditions of the
extended atmosphere in pulsating stars.

\section{Conclusions}

 We reported the time variation in H$_2$O bands in four oxygen-rich
Mira variables.
The near infrared water-vapour bands at 2.5--3.95\,$\mu$m follow
the periodical variation.
Emission features are seen at $\sim$3.5--3.95\,$\mu$m around maximum
while absorption features are detected around minimum.
The spectra are well fitted with `slab' models,
which consist of two H$_2$O layers (hot layer and cool layer).
The radius of the hot layer varies from $\sim$1\,$R_*$ to
$\sim$2\,$R_*$ during visual minimum and maximum.

 The periodical variation in the features arising from the extended
atmosphere suggests that the structure of the outer atmosphere is varying
with the pulsation.
The pulsation produces a shock in the atmosphere,
and the hot H$_2$O layer traces the high H$_2$O density caused by the shock.
The high H$_2$O density shell expands from inside of 
the extended atmosphere to outwards
from minimum to maximum.
In our analysis, $R_*$ is measured as the relative number of 
the stellar radius.
Considering the variation in the radius of the star,
the high density shell is by a factor of 1.5--2.0 more extended at maximum
than at minimum.

 Due to large optical depth of H$_2$O bands,
the spectra in the 2.5--4.0\,$\mu$m region
are dominated by the H$_2$O in the extended atmosphere.
The flux variation in the L'-band is primarily determined by the radial motion
of optically thick H$_2$O layer.

\begin{acknowledgements}
M.M. thanks people at the University of Amsterdam, especially
Prof.\,L.B.F.M.\ Waters
for the hospitality during her stay. 
We acknowledge discussions with Prof. T.\ de~Jong.
Comments with Drs.\,A.A.\ Zijlstra and J.M.\ Winters 
improved this paper.
We appreciate careful reading and comments of the anonymous referee.
M.M. was the Research Fellow of the 
Japan Society for the Promotion of Science for the Young Scientists.
I.Y. acknowledges support by Grant-in-Aid for Encouragement of
Young Scientists (No. 13740131) from Japan Society for the Promotion
of Science.
\end{acknowledgements}

\end{document}